\documentclass[runningheads,a4paper]{llncs}

\usepackage{hyperref}
\usepackage{color,soul}

\usepackage{caption}
\usepackage{subfig}

\usepackage{amssymb}
\usepackage{graphicx}

\usepackage{url}
\urldef{\mailsa}\path|{sourla, zaro}@ceid.upatras.gr|
\urldef{\mailsb}\path|sioutas@ionio.gr|
\urldef{\mailsc}\path|tsichlas@csd.auth.gr|
\newcommand{\keywords}[1]{\par\addvspace\baselineskip
\noindent\keywordname\enspace\ignorespaces#1}

\begin{document}

\spnewtheorem{invariant}{Invariant}{\itshape}{\rmfamily}
\spnewtheorem{algorithm1}{Algorithm}{\bfseries}{\rmfamily}

\mainmatter  

\title{D3-Tree: A Dynamic Distributed Deterministic Load - Balancer for decentralized tree structures}

\titlerunning{D3-Tree: A Dynamic Distributed Deterministic Load - Balancer}

%
%
\author{Efrosini Sourla\inst{1}
\and Spyros Sioutas\inst{2}\and Kostas Tsichlas\inst{3}
\and Christos Zaroliagis\inst{1}
}
\authorrunning{E. Sourla, S. Sioutas and K. Tsichlas and C. Zaroliagis}

\institute{Department of Computer Engineering and Informatics,\\
University of Patras, 26500 Patras, Greece\\
\mailsa\\
\and Department of Informatics, Ionian University, 49100 Corfu, Greece\\
\mailsb\\
\and Department of Informatics, Aristotle University of Thessaloniki,\\
54124 Thessaloniki, Greece\\
\mailsc\\
}

%
%

\toctitle{D3-Tree: A Dynamic Distributed Deterministic Load - Balancer for decentralized hierarchical tree structures}
\tocauthor{E. Sourla, S. Sioutas and K. Tsichlas and C. Zaroliagis}
\maketitle

\centerline{\today}

\begin{abstract}
In this work, we propose D$^3$-Tree, a dynamic distributed deterministic structure for data management in decentralized networks. We present in brief the theoretical algorithmic analysis, in which our proposed structure is based on, and we describe thoroughly the key aspects of the implementation. 
Conducting experiments, we verify that the implemented structure outperforms other well-known hierarchical tree-based structures, since it provides better complexities regarding load-balancing operations. More specifically, the structure achieves an $O(\log{N})$ amortized bound ($N$ is the number of nodes present in the network), using an efficient deterministic load-balancing mechanism, which is general enough to be applied to other hierarchical tree-based structures. Moreover, we investigate the structure's fault tolerance, which hasn't been sufficiently tackled in previous work, both theoretically and through rigorous experimentation. We prove that D$^3$-Tree is highly fault tolerant, since, even for massive node failures, it achieves a significant success rate in element queries. Afterwards we go one step further, in order to achieve sub-logarithmic complexity and propose the ART$^+$ structure (Autonomous Range Tree), exploiting the excellent performance of D$^3$-Tree. ART$^+$ achieves an $O(\log_b^2{\log{N}})$ communication cost for query and update operations ($b$ is a double-exponentially power of $2$ and $N$ is the total number of peers). Moreover, ART$^+$ is a fully dynamic and fault-tolerant structure, which supports the join/leave node operations in $O(\log{\log{N}})$ expected w.h.p number of hops and performs load-balancing in $O(\log{\log{N}})$ amortized cost.
\end{abstract}


\section{Introduction}
\label{sec:intro}
Decentralized systems and in particular Peer-to-Peer (P2P) networks, have generated a lot of interest worldwide among the computer networking community. Although they have actually existed for many years, they have become very popular nowadays and are promoted as the future of Internet networking. They are widely used for sharing resources and store very large data sets, using systems of small computers instead of large costly servers.

According to \cite{s02:p2pdef}, a P2P network "is a type of decentralized and distributed network architecture, in which, individual nodes in the network (called "peers") act as both suppliers and consumers of resources, in contrast to centralized client-server model where client nodes request access to resources provided by central servers. In a peer-to-peer network, tasks (such as searching for files or streaming audio/video) are shared amongst multiple interconnected peers, who each make a portion of their resources (such as processing power, disk storage or network bandwidth) directly available to other network participants, without the need for centralized coordination by servers".

In P2P networks, data are stored at the nodes (or peers) and the most crucial operations are data search and data updates. A P2P network is represented by a graph, a logical \textit{overlay network}, where its nodes correspond to the network nodes, while its edges may not correspond to existing communication links, but to communication paths.
We assume constant size messages between nodes through links and asynchronous communication. It is assumed that the network provides an upper bound on the time needed for a node to send a message and receive an acknowledgement. In this way, the network provides a mechanism to identify communication problems, which may refer to communication links or nodes that are down.
The complexity (cost) of an operation is measured in terms of the number of messages issued during its execution (internal computations at nodes are considered insignificant).

With respect to its \textit{structure}, the overlay supports the operations \textit{Join} (of a new node $v$; $v$ communicates with an existing node $u$ in order to be inserted into the overlay), and \textit{Departure} (of an existing node $u$; $u$ leaves the overlay announcing its intent to other nodes of the overlay). Moreover, the overlay implements an \textit{indexing scheme} for the stored data, supporting the operations \textit{Insert} a new element, \textit{Delete} an existing element, \textit{Search} for an element and \textit{Range Query} for elements in a specific range.

Range query processing in decentralized network environments is a notoriously difficult problem to solve efficiently and scalably.
In cloud infrastructures, a most significant and apparent requirement is the monitoring of thousands of computer nodes, which often requires support for range queries: consider range queries issued in order to identify under-utilized nodes so as to assign them more tasks, or to identify overloaded nodes so as to avoid bottlenecks in the cloud. For example, we wish to execute range queries such as:
\begin{verbatim}
SELECT NodeID
FROM CloudNodes
WHERE Low < utilization < High
\end{verbatim}
or both single and range queries such as:
\begin{verbatim}
SELECT NodeID
FROM CloudNodes
WHERE Low < utilization < High AND os = UNIX
\end{verbatim}

Moreover, in cloud infrastructures that support social network services like Facebook, user profiles are stored distributed in several nodes and we wish to retrieve user activity information, executing range queries such as:
\begin{verbatim}
SELECT COUNT(userID)
FROM CloudNodes
WHERE 3/1/2015 < time < 3/31/2015 AND userID = 123456
      AND NodeID IN Facebook
\end{verbatim}

An acceptable solution for processing range queries in such large-scale decentralized environments must scale in terms of the number of nodes as well as in terms of the number of data items stored.
For very large volume data (trillions of data items at millions of nodes) the classic logarithmic complexity offered by solutions in literature, is still too expensive for single and range queries.
Further, all available solutions incur large overheads with respect to other critical operations, such as join/leave of nodes, and insertion/deletion of items.
Our aim in this work is to provide a solution that is comprehensive and outperforms related work with respect to all major operations, such as search, join/leave, insert/delete and load-balancing and to the required routing state that must be maintained in order to support these operations. In particular, our ultimate goal is to achieve a sub-logarithmic complexity for all the above operations.

In this work, we focus on hierarchical tree-based overlay networks that support directly range and more complex queries. We introduce a much promising deterministic decentralized structure for distributed data, called D$^3$-Tree.
Through experiments, we verify that the implemented structure outperforms other well-known tree-based structures, since it provides better complexities regarding load-balancing operations. More specifically, the structure achieves an amortized bound of $O(\log{N})$ ($N$ is the number of nodes present in the network), using an efficient deterministic weight-based load-balancing mechanism, which is general enough to be applied to other hierarchical tree-based structures. Moreover, our structure achieves an $O(\log{N})$ search performance. Last but not least, we investigate the structure's fault tolerance, which has not been sufficiently tackled in its predecessor\cite{bstz14}, both theoretically and through rigorous experimentation, proving that D$^3$-Tree is highly fault-tolerant.

Afterwards we go one step further, in order to achieve sub-logarithmic complexity and propose the ART$^+$ structure, exploiting the excellent performance of D$^3$-Tree.
The outer level of ART$^+$ is an ART\footnote{Autonomous Range Tree} structure\cite{stpstm2012:art}, built by grouping clusters of peers, whose communication cost of query and update operations is $O(\log_b^2{\log{N}})$ hops, where the base $b$ is a double-exponentially power of two and $N$ is the total number of nodes. Moreover, ART is a fully dynamic and fault-tolerant structure, which supports the join/leave node operations in $O(\log{\log{N}})$ expected w.h.p number of hops and performs load-balancing in $O(\log{\log{N}})$ amortized cost.
Each cluster-peer of ART$^+$ is organized as a D$^3$-Tree.

The rest of this paper is organized as follows: Previous work is presented in Section \ref{sec:previous}.
The weight-based mechanism used in D$^3$-Tree is briefly described in Section \ref{sec:mechanism}.
Section \ref{sec:d3tree} presents our proposed structure by describing the theoretical background and discussing the enhancements and implementation aspects.
In Section \ref{sec:art+} we present the ART$^+$ structure which exploits the performance of D$^3$-Tree.
Section \ref{sec:experiments} hosts the performance evaluation for both structures.
The paper concludes in Section \ref{sec:conclusions}.


\section{Related Work}
\label{sec:previous}
Extensive work has been done to support search and update techniques in distributed data, which are crucial operations in P2P networks. In this section we discuss the existing solutions referring to their contribution and weaknesses, emphasizing in hierarchical tree-based structures. 

Existing structured P2P systems can be classified into two broad categories: Distributed Hash Table (DHT)-based systems and tree-based systems.
Examples of the former, which constitute the majority, include Chord, CAN, Pastry, Symphony, Tapestry \cite{ov2011} and P-Ring \cite{clmgs2011:pring}. In general, DHT-based systems support exact match queries well and use (successfully) probabilistic methods to distribute the workload among nodes equally.
Since hashing destroys the ordering on keys, DHT-based systems typically do not possess the functionality to support straightforwardly range queries, or more complex queries based on data ordering (e.g., nearest-neighbour and string prefix queries). Some efforts towards addressing range queries have been made in \cite{GAA2003,SGAA2004}, getting however approximate answers and also making exact searching highly inefficient.
The most recent effort towards range queries is the P-Ring \cite{clmgs2011:pring}. P-Ring is fully distributed and fault-tolerant, provides load-balancing and supports both exact match and range queries, achieving $O(\log_d{N} + k)$ range search performance in average case ($N$ is the number of peers, $d$ is the \textit{order} of the ring and $k$ is the answer size) and $O(d \cdot \log_d{N} + k)$ in worst case.

Tree-based systems are based on hierarchical structures. They support range queries more naturally and efficiently as well as a wider range of operations, since they maintain the ordering of data. On the other hand, they lack the simplicity of DHT-based systems, and they do not always guarantee data locality and load balancing in the whole system. 
Important examples of such systems include Family Trees \cite{ov2011}, BATON \cite{jov05}, BATON$^*$\cite{jotvz06} and Skip List-based schemes like Skip Graphs (SG), NoN SG, SkipNet (SN), Deterministic SN, Bucket SG, Skip Webs, Rainbow Skip Graphs (RSG) and Strong RSG \cite{ov2011} that use randomized techniques to create and maintain the hierarchical structure.

Emphasis should be given to the fact that w.r.t. load-balancing, the solutions provided in the literature are either heuristics, or provide expected bounds under certain assumptions, or amortized bounds but at the expense of increasing the memory size per node. In particular, in BATON \cite{jov05}, a decentralized overlay is provided with load-balancing based on data migration. However, their $O(\log{N})$ amortized bound ($N$ is the number of nodes in the network) is valid only subject to a probabilistic assumption about the number of nodes taking part in the data migration process, and thus it is in fact an amortized expected bound. Moreover, its successor BATON$^*$, exploits the advantages of higher \textit{fanout} (number of children per node), to achieve reduced search cost of $O(\log_m{N})$, where $m$ is the \textit{fanout}. However, the higher \textit{fanout} leads to larger update and load-balancing cost of $O(m \cdot \log_m{N})$.
On the other hand, in DHT systems, P-Ring \cite{clmgs2011:pring} maintains a load imbalance factor of at most $2+\epsilon$ in a stable system, for any given constant $\epsilon > 0$, and has a \textit{stabilization process} for fixing inconsistencies caused by peer failures and updates, achieving an $O(d \cdot \log_d{N})$ performance.

As far as network's fault tolerance is concerned, 
P-Ring \cite{clmgs2011:pring} is considered highly fault-tolerant, using the Chord's Fault Tolerant Algorithms \cite{Stoica:chord}. BATON \cite{jov05} maintains vertical and horizontal routing information not only for efficient search, but to offer a large number of alternative paths between two nodes. In its successor BATON$^*$ \cite{jotvz06}, fault tolerance is greatly improved due to higher \textit{fanout}. When $fanout = 2$, approximately 25$\%$ of nodes must fail before the structure becomes partitioned, while increasing the fanout up to 10 leads to increasing fault tolerance ($60\%$ of failed nodes partition the structure).
A comparison of the aforementioned structures and our proposed structure is given in Table \ref{table:comp}.

\begin{table}
\centering
\caption{Comparison of P-Ring, BATON$^*$, D$^3$-Tree, ART and ART$^+$.}
\setlength{\tabcolsep}{5pt}
\begin{tabular}{ l  l  p{1in}  p{0.9in}  p{1in} }
\hline\noalign{\smallskip}
Structures & Search key & Insert/Delete key (load-balancing) & Max. size of routing table & Join/Depart peer (updating routing tables)\\
\noalign{\smallskip}
\hline
\noalign{\smallskip}
P-Ring & $O(\log_d{N})$ & $\widetilde{O}(d \cdot \log_d{N})$ & $O(\log{N})$ & $\widetilde{O}(d \cdot \log_d{N})$ \\ 
BATON & $O(\log{N})$ & $\overline{O}(\log{N})$ & $O(\log{N})$ & $\overline{O}(\log{N})$ \\ 
BATON$^*$ & $O(\log_m{N})$ & $\overline{O}(m \cdot \log_m{N})$ & $O(m \cdot \log_m{N})$ & $\overline{O}(m \cdot \log_m{N})$ \\ 
D$^3$-Tree & $O(\log{N})$ & $\widetilde{O}(\log{N})$ & $O(\log{N})$ & $\widetilde{O}(\log{N})$ \\ 
ART & $\widehat{O}(\log_{b}^2{\log{N}})$ & $\overline{O}(m \cdot \log_m{\log{N}})$ & $O(N^{1/4} / \log^c{N})$ & $\widehat{O}(m \cdot \log_m{\log{N}})$ \\ 
ART$^+$ & $\widehat{O}(\log_{b}^2{\log{N}})$ & $\widetilde{O}(\log{\log{N}})$ & $O(N^{1/4} / \log^c{N})$ & $\widehat{O}(\log{\log{N}})$ \\ 
\multicolumn{5}{l}{ } \\
\multicolumn{5}{p{4.7in}}{Legend: $N$: number of peers, $d$: order of ring, $m$: fanout, $c > 0$, $b$: double-exponentially power of 2, $\widehat{O}$: expected bound, $\widetilde{O}$: amortized bound, $\overline{O}$: expected amortized bound.} \\ \hline
\end{tabular}
\label{table:comp}
\end{table}


\section{A Weight-Based Load-Balancer}
\label{sec:mechanism}

In this section, we describe a weight-based load-balancing mechanism which is an efficient solution for element updates in hierarchical tree-based structures. All definitions used in this section and throughout this paper, have been compiled in Table \ref{table:def}. The main idea of this mechanism is the almost equal distribution of elements among nodes by making use of \emph{weights}, a metric which shows how uneven is the load among nodes. When the load is uneven, then a data migration process is initiated to equally distribute the elements. The method has two steps; (a) first, it provides efficient and local update of weight information in a tree when elements are added or removed at the leaves, using \textit{virtual weights}, and (b) it provides an efficient load-balancing mechanism which is activated when necessary.

\begin{table}
\centering
\caption{Symbols and Definitions}
\setlength{\tabcolsep}{5pt}
\begin{tabular}{c p{8.5cm}}
\hline\noalign{\smallskip}
Symbol&Definition\\
\noalign{\smallskip}
\hline
\noalign{\smallskip}
$w(v)$ : weight of $v$ & number of elements stored in the subtree of $v$(including $v$)\\ 
$e(v)$ & number of elements residing in a node $v$\\ 
$|v|$ : size of $v$ & number of nodes of the subtree of $v$ (including $v$)\\ 
$d(v)$ : density of $v$ & $d(v)=\frac{w(v)}{|v|}$ represents the mean number of elements per node in the subtree of $v$\\ 
$c(p,q)$ : criticality & $c(p,q)=\frac{d(p)}{d(q)}$ represents the difference in densities between brothers $p$ and $q$\\ 
$nc_v$ : node criticality & $nc_v=\frac{|w|}{|v|}$ represents the difference in size between a node $v$ and its left child $w$\\ \hline
\end{tabular}
\label{table:def}
\end{table}

More specifically, when an element is added/removed to/from a leaf $u$ in a tree structure $\mathcal{T}$, the weights on the path from $u$ to the root must be updated. This is a costly operation, when it is performed on every element update. Instead of updating weights every time, a new metric is defined, \textit{virtual weight} $W(v)$. Assume that node $v$ lies at height $h$ and its children are $v_{1},v_{2},\ldots,v_{s}$ at height $h-1$. $W(v)$ of $v$ is defined as the weight stored in node $v$. In particular, for node $v$ the following invariants are maintained:

\begin{invariant} \label{inv:1}
$W(v)>e(v)+(1-\epsilon_{h})\left(\sum_{i=1}^{s}{W(v_{i})}\right)$
\end{invariant}
\begin{invariant} \label{inv:2}
$W(v)<e(v)+(1+\epsilon'_{h})\left(\sum_{i=1}^{s}{W(v_{i})}\right)$
\end{invariant}

where $\epsilon_{h} = \epsilon'_{h} = \frac{1}{h^{2}}$. The constants $\epsilon_{h}$ and $\epsilon'_{h}$ are chosen such that for all nodes the virtual weight will be within a constant factor $c>1$ of the real weight, i.e., $\frac{1}{c}\cdot w(v) < W(v) < c\cdot w(v)$.

When an update takes place at leaf $u$, the mechanism traverses the path from $u$ to the root updating the weights by $\pm 1$, until node $z$ is found for which Invariants~\ref{inv:1} and \ref{inv:2} hold. Let $v$ be its child for which either Invariant~\ref{inv:1} or \ref{inv:2} does not hold on this path. All weights on the path from $u$ to $v$ are recomputed; for each node $z$ on this path, its weight information is updated by taking the sum of the weights of its children plus the number of elements that $z$ carries. 

Another Invariant which is maintained and is crucial for the load-balancing mechanism, involves the criticality $c(p,q) = \frac{d(p)}{d(q)}$ of two brother nodes $p$ and $q$ (representing their difference in densities). The invariant guarantees that there will not be large differences between densities:

\begin{invariant} \label{inv:3}
For two brothers $p$ and $q$, it holds that $\frac{1}{c} \leq c(p,q) \leq c, 1< c \leq 2$
\end{invariant}

For example, choosing $c=2$ we get that the density of any node can be at most twice or half of that of its brother.

When an update takes place at leaf $u$, weights are updated as described above. Then, the load-balancing mechanism redistributes the elements among leaves when the load between leaves is not distributed equally enough. In particular, starting from $u$, the highest ancestor $w$ is located that is unbalanced w.r.t. his brother $z$, meaning that Invariant~\ref{inv:3} is violated. Finally, the elements in the subtree of their father $v$ are redistributed uniformly so that the density of the brothers becomes equal; this procedure is henceforth called \textit{redistribution} of node $v$.

The weight-based mechanism is slightly modified to be applied also in node updates. \textit{Virtual size} $S(v)$ is defined and the same mechanism is applied using invariants similar to Invariants \ref{inv:1} and \ref{inv:2}.
Moreover, a new invariant is defined, involving node criticality.

\begin{invariant} \label{inv:nodecriticality}
 The node criticality of all nodes is in the range $\left[\frac{1}{4},\frac{3}{4}\right]$.
\end{invariant}

Invariant~\ref{inv:nodecriticality} implies that the number of nodes in the left subtree of a node $v$ is at least half and at most twice the corresponding number of its right subtree.

The weight-based mechanism described above (and its slight modification) achieves an $O(1)$ amortized cost for weight update and $O(\log{N})$ amortized cost for redistribution. It is general enough to be applied to other hierarchical tree-based structures and was proposed and thoroughly described in \cite{bstz14}. 

\section{The D$^3$-Tree}
\label{sec:d3tree}
In this section, we present our proposed structure, D$^3$-Tree, which introduces many enhancements over the solutions in literature and its predecessor \cite{bstz14}.
In general, a D$^3$-Tree structure with $N$ nodes and $n$ data elements residing on them achieves: (i) $O(\log{N})$ space per node; (ii) deterministic $O(\log{N})$ searching cost; (iii) deterministic amortized $O(\log{N})$ update cost both for element updates and for node joins and departures; (iv) deterministic amortized $O(\log{N})$ bound for load-balancing.

\subsection{The Structure}
\label{subsec:structure}

Let $N$ be the number of nodes present in the network and let $n$ denote the size of data ($N \ll n$). The structure consists of two levels. The upper level is a Perfect Binary Tree (PBT) of height $O(\log{N})$. The leaves of this tree are \textit{representatives} of the buckets that constitute the lower level of the D$^3$-Tree. Each bucket is a set of $O(\log{N})$ nodes which are structured as a doubly linked list.
The number of nodes of the PBT is not connected by any means to the number of elements stored in the structure.
The structure supports the operations of node join and node departure, while at the same time it tackles failures of nodes whenever these are discovered.
Each node $v$ of the D$^3$-Tree maintains an additional set of links to other nodes apart from the standard links which form the tree:

\begin{enumerate}
 \item Links to its father and its children.
 \item Links to its adjacent nodes based on an in-order traversal of the tree.
 \item Links to nodes at the same level as $v$. 
  The links are distributed in exponential steps; the first link points to a node (if there is one) $2^{0}$ positions to the left (right), the second $2^{1}$ positions to the left (right), and the $i$-th link $2^{i-1}$ positions to the left (right). These links constitute the \textit{routing table} of $v$ and require $O(\log{N})$ space per node.
 \item Links to leftmost and rightmost leaf of its subtree. These links accelerate the search process and contribute to the structure's fault tolerance when a considerable number of nodes fail.
  \item For leaf nodes only, links to the buckets of the nodes in their routing tables. The first link points to a bucket $2^0$ positions left (right), the second $2^1$ positions to the left (right) and the $i$-th link $2^{i-1}$ positions to the left (right). These links require $O(\log{N})$ space per node and keep the structure fault tolerant, since each bucket has multiple links to the PBT.
\end{enumerate}

The next lemma captures some important properties of the routing tables.

\begin{lemma}\label{lem:properties}
(i) If a node $v$ contains a link to node $u$ in its routing table, then the parent of $v$ also contains a link to the parent of $u$, unless $u$ and $v$ have the same father.
(ii) If a node $v$ contains a link to node $u$ in its routing table, then the left (right) sibling of $v$ also contains a link to the left (right) sibling of $u$, unless there are no such nodes.
(iii) Every non-leaf node has two adjacent nodes in the in-order traversal, which are leaves.
\end{lemma}

Regarding the index structure of the D$^3$-Tree, the range of all values stored in it is partitioned into sub-ranges each one of which is assigned to a node of the overlay. An internal node $v$ with range $[x_v,x'_v]$ may have a left child $u$ and a right child $w$ with ranges $[x_u,x'_u]$ and $[x_w,x'_w]$ respectively such that $x_u<x_u'<x_v<x_v'<x_w<x_w'$. Ranges are dynamic in the sense that they depend on the values maintained by the node.

\subsection{Implementation Aspects}
\label{sec:implementation}

During the implementation process, many of the theoretical aspects of the proposed structure were put under the microscope. The transition from theory to practice is full of challenges, leading to issues that either can be solved or have to be simplified. Below we describe the key features of the D$^3$-Tree simulator and the algorithmic steps behind the operations it supports.

A key feature of our proposed structure, thanks to which the high performance is achieved, is the weight-based mechanism \cite{bstz14}, used for node redistribution after node updates and data load-balancing after element updates. The main idea is the almost equal distribution of elements among nodes, using \textit{weights}, a metric which shows how uneven is the load among nodes. The mechanism, which was described in Section \ref{sec:mechanism} lazily updates the weight information on nodes, so load-balancing is performed only when it is absolutely necessary.

Another key feature is the enhanced search mechanism, in case of node failures. Nodes perform a number of effective horizontal contacts to other nodes of the same level and take into account several alternative paths, in order the operation to be successful, even when a considerable number of nodes fails.
Last but not least, the structure is highly fault tolerant, since is supports a procedure of node withdrawal, when a node is found unreachable, regardless of its position(internal node, leaf, bucket node). The success of the last two operations is due to the small number of additional links a node maintains, through which it can reconstruct the routing table of a fallen node.

\subsection{Node Joins and Departures}
\label{subsec:nodes}
\subsubsection{Handling node updates.}
When a node $z$ makes a join request (Alg. \ref{alg:join}) to $v$, $v$ forwards the request to an adjacent leaf $u$. If $v$ is a binary node, the request is forwarded to the left adjacent node, w.r.t. the in-order traversal, which is definitely leaf (unless $v$ is a leaf itself). In case $v$ is a bucket node, the request is forwarded to the bucket representative, which is leaf. Then, node $z$ is added to the doubly linked list of the bucket represented by $u$. In node joins, we make the simplification that the new node is clear of elements.
It could be entered anywhere in the bucket, as the first, the last or an intermediate node, but we prefer to place it after the most loaded node of the bucket. Thus, the load is shared and the new node stores half of the elements of the most loaded node. 
Finally, the left and right adjacents of the newly inserted node update their links to previous and next node and $z$ creates new links to those nodes.

\begin{algorithm1}
Join (Node newNode)
\label{alg:join}
\begin{verbatim}
   REQUIRE initialNode
   IF initialNode is BucketNode THEN
      initialNode.Representative.Join(newNode);
   ELSE IF initialNode is InternalBinaryNode THEN
      initialNode.LeftInOrderAdjacent.Join(newNode);
   ELSE IF initialNode is Leaf THEN
      mostLoadedNode = initialNode.FindMostLoaded();
      mostLoadedNode = AcceptAsAdjacent(newNode);
      SplitData(mostLoadedNode, newNode);
   END IF
\end{verbatim}
\end{algorithm1}

When a node $v$ leaves the network, it is replaced by an existing node, so as to preserve the in-order adjacency. All navigation data are copied from the departing node $v$ to the replacement node, along with the elements of $v$. If $v$ is an internal binary node, then it is replaced by its right adjacent node, which is a leaf and which in turn is replaced by the first node $z$ in its bucket. If $v$ is a leaf, then it is directly replaced by $z$. 
If the departing node belongs to a bucket no replacement takes place, but simply all stored elements of $v$ are copied to the previous node of $v$ (or to the next node if the previous does not exist).
Then $v$ is free to depart.

After a node join or departure, the modified weight-based mechanism is activated and updates the sizes by $\pm 1$ on the path from leaf $u$ to the root (Alg. \ref{alg:update_size}), as long as the defined Invariants \ref{inv:1} or \ref{inv:2} do not hold. When the first node $w$ is accessed for which Invariants \ref{inv:1} and \ref{inv:2} hold, the nodes in the subtree of its child $q$ in the path, have their sizes recomputed (Alg. \ref{alg:comp_size}). Afterwards, the mechanism traverses the path from leaf $u$ to the root, in order to find the first node (if such a node exists) for which Invariant \ref{inv:nodecriticality} is violated (Alg. \ref{alg:node_criticality}) and performs a redistribution in its subtree.

\begin{algorithm1}
UpdateVirtualSize (BinaryNode leaf, NodeUpdate operation)
\label{alg:update_size}
\begin{verbatim}
   currentNode = leaf;
   WHILE currentNode.Father != null DO
      IF operation == NodeJoin THEN
         currentNode.Father.Size++;
      ELSE IF operation == NodeDeparture THEN
         currentNode.Father.Size--;
      END IF
      /* if Invariant 1 OR Invariant 2 do not hold */
      IF currentNode.Father.Size <= minSize OR 
      currentNode.Father.Size >= maxSize THEN
         currentNode = currentNode.Father;
      ELSE
         RETURN currentNode;
      END IF
   END WHILE
   RETURN Root;
\end{verbatim}
\end{algorithm1}

\begin{algorithm1}
ComputeSizeInSubtree ()
\label{alg:comp_size}
\begin{verbatim}
   REQUIRE node
   IF node IS Leaf THEN
      node.Size = node.Bucket.Size;
   ELSE
      node.LeftChild.ComputeSizeInSubtree();
      node.RightChild.ComputeSizeInSubtree();
      node.Size = node.LeftChild.Size + node.RightChild.Size;
      node.minSize = (1 - $e_h$) * node.Size;
      node.maxSize = (1 + $e_h$) * node.Size;
   END IF
\end{verbatim}
\end{algorithm1}

\begin{algorithm1}
CheckNodeCriticality(BinaryNode leaf)
\label{alg:node_criticality}
\begin{verbatim}
   currentNode = leaf.Father;
   WHILE currentNode != null DO
      IF currentNode.NodeCriticality IN [minValue, maxValue] THEN
         currentNode = currentNode.Father;
      ELSE
         RETURN currentNode;
      END IF
   END WHILE
   RETURN null; /* redistribution isn't necessary */
\end{verbatim}
\end{algorithm1}

\subsubsection{Node Redistribution.} \label{subsec:redistribution}
The redistribution guarantees that if there are $z$ nodes in total in the $y$ buckets of the subtree of $v$, then after the redistribution each bucket maintains either $\lfloor z/y \rfloor$ or $\lfloor z/y \rfloor +1$ nodes. The redistribution cost is $O(\log{N})$, which is verified through experiments.

The redistribution in the subtree of $v$ is carried out as follows (Alg. \ref{alg:redistribution}). We assume that the subtree of $v$ at height $h$ has $K$ buckets. A traversal of all buckets is performed in order to determine the exact value of $|v|$ (number of nodes in the buckets of the subtree of $v$). Then, the first $k$ buckets, will contain $\left\lfloor \frac{|v|}{2^{h}}\right\rfloor+1$ nodes after redistribution, where $k = |v|\bmod{2^h}$. The remaining $K - k$ buckets will contain $\left\lfloor \frac{|v|}{2^{h}}\right\rfloor$ nodes. The redistribution starts from the rightmost bucket $b$ and it is performed in an in-order fashion so that elements in the nodes are not affected.

We assume that $b$ has $q$ extra nodes which must be transferred to other buckets. Bucket $b$ maintains a link \textit{dest} to the next bucket $b'$ on the left, in which $q$ extra nodes should be put. The $q$ extra nodes are removed from $b$ and are added to $b'$. The crucial step in this procedure is that during the transfer of nodes, internal nodes of PBT are also updated, since the in-order traversal must remain untouched. More specifically, the representative $z$ of $b$ as well as the its left in-order adjacent $w$ are replaced by nodes of bucket $b$ and then, $z$ and $w$ along with the remaining $q-2$ nodes of $b$ are added to the tail of bucket $b'$. Afterwards, the horizontal and vertical links for the replaced binary nodes are updated, as well as for nodes that point to them. Finally, bucket $b$ informs $b'$ to take over and the same procedure applies again with $b'$ as the source bucket.

The case where $q$ nodes must be transferred to bucket $b$ from bucket $b'$ is completely symmetric. In general, $q$ nodes are removed from the tail of bucket $b'$, two of them replace nodes $w$ and $z$ and the remaining nodes are added to the head of $b$. The intriguing part comes when $b'$ contains less nodes than the $q$ that $b$ needs. In this case, $b$ has to find the remaining nodes in the buckets on the left, so \textit{dest} travels towards the leftmost bucket of the subtree, until $q \leq \sum_{i=1}^{s}|{b_{i}|}$, where $|b_{i}|$ is the size of the $i-th$ bucket on the left. Then, nodes of $b_{s}$ move to $b_{s-1}$, nodes from $b_{s-1}$ are transfered to $b_{s-2}$ and so on, until \textit{dest} goes backwards to $b'$ and $q$ nodes are moved from $b'$ into $b$.

The redistribution cost is $O(\log{N})$ and is verified through experiments presented analytically in this work.

\begin{algorithm1}
RedistributeSubtree()
\label{alg:redistribution}
\begin{verbatim}
   REQUIRE BinaryNode node
   node.ComputeSizeInSubtree();
   newBucketSize =  node.Size / 2^h;
   leftMostLeaf = node.LeftMostLeaf;
   currentNode = node.RightMostLeaf;
   WHILE currentNode  != leftMostLeaf DO
      currentBucket = currentNode.Bucket;
      destinationBucket = currentNode.LeftAdjacents[0].Bucket;
      IF currentBucket.Size > newBucketSize THEN
         nodesToMove = currentBucket.Size - newBucketSize;
         FOR i=0; i < nodesToMove; i++ DO
            Replace(currentNode, currentBucket.FirstNode)
            Replace(currentNode.LeftAdjacent, currentBucket.Representative);
            destinationBucket.InsertLast(currentNode.LeftAdjacent);
            UpdateAllLinks();
         END FOR
      ELSE IF currentBucket.Size < newBucketSize THEN
         nodesToMove = newBucketSize - currentBucket.Size;
         tempDestNode = currentNode.LeftAdjacents[0];
         WHILE tempDestNode != leftMostLeaf AND 
         availableNodes < nodesToMove DO
            availableNodes += tempDestNode.Size;
            tempDestNode = tempDestNode.LeftAdjacents[0];
         END WHILE
         tempNodesToMove = tempDestNode.Size - 
         (totalNodesAvailable - nodesToMove);
         WHILE tempDestNode != currentNode DO
            /* ... similar to previous case ...*/
            /* go back to the right, moving nodes */
         END WHILE
      END IF
      currentNode = currentNode.LeftAdjacents[0];
   END WHILE
\end{verbatim}
\end{algorithm1}

\subsubsection{Extension - Contraction.}
\begin{figure}
\centering
\includegraphics[width=4in]{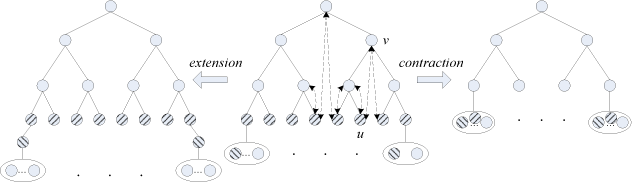}
\caption{The initial D$^3$-Tree structure (middle) and the operations of extension (left) and contraction (right).}
\label{fig:ops}
\end{figure}

Throughout joins and departures of nodes, the size of buckets can increase undesirably or can decrease so much that some buckets may become empty. Either situations violate the $D^3$-Tree principle for bucket size $|b|$:
\begin{equation}
\label{eq:1}
 a_{1}\log{N} \leq |b| \leq a_{2}\log{N}
\end{equation}
where $0 < a_{1} \leq a_{2}$. The structure guarantees that (\ref{eq:1}) is always true, by employing two operations on the PBT, \textit{extension} and \textit{contraction} (Figure \ref{fig:ops}).
These operations are activated when a redistribution occurs at the root of the PBT and they add one level to the PBT when $|b| > a_{2}\log{N}$ or delete one level from the PBT when $|b| < a_{1}\log{N}$.

The extension (Alg. \ref{alg:extension}) is carried out as follows: the last level of the PBT is affected and a new level of leaves, as well as, a new set of buckets are created, using nodes from the old buckets. In particular, each leaf $u$ and its bucket $B$ with nodes $b_{1}, b_{2}, ..., b_{s}$ are replaced by a 2-level binary subtree $T_{2}$ preserving, at the same time, the in-order adjacency. Thus, the left leaf of $T_{2}$ is the old leaf $u$, the root of $T_{2}$ is the $i-th$ node of $b$, where $i = \left\lceil \frac{|b|}{2} \right\rceil$, the bucket $B_{1}$ of $u$ contains nodes $b_{1}, ..., b_{i-1}$, the right leaf of $T_{2}$ is the $(i+1)-th$  node of $b$ and its bucket $B_{2}$ contains the remaining nodes of old bucket $B$, $b_{i+2},...,b_{s}$. During the process, all in-order adjacency links and father-child links are updated. After the extension has been carried out, leaves and nodes in height $h = 1$ reconstruct their routing tables and all binary nodes update the link to the rightmost leaf of their subtree\footnote{the leftmost leaf is not affected}.

\begin{algorithm1}
PerformExtension()
\label{alg:extension}
\begin{verbatim}
   REQUIRE BinaryNode Root
   currentLeaf = Root.LeftMostLeaf;
   WHILE currentLeaf != null DO
      grandFather = currentLeaf.Father;
      oldBucket = currentLeaf.Bucket;
      i = Math.Ceiling(oldBucket.Size/2);
      B1 = oldBucket.RemoveNodes(i);
      newFather = oldBucket.RemoveFirstNode();
      rightLeaf = oldBucket.RemoveFirstNode();
      B2 = oldBucket.RemoveRemainingNodes();
      grandFather.ReplaceChild(currentLeaf,newFather);
      newFather.SetChildren(currentLeaf, rightLeaf);
      UpdateInOrderAdjacencyLinks();
      currentLeaf = currentLeaf.RightAdjacents[0];
   END WHILE
   UpdateAllLinks();
   Root.ComputeSizeInSubtree();
\end{verbatim}
\end{algorithm1}

\begin{algorithm1}
PerformContraction()
\label{alg:contraction}
\begin{verbatim}
   REQUIRE BinaryNode Root
   currentNode = Root.LeftMostLeaf.Father;
   WHILE currentNode != null DO
      currentLeaf = currentNode.LeftChild;
      newFather = currentNode.Father;
      newBucket = currentLeaf.Bucket;
      newBucket.Add(currentLeaf.Father);
      newBucket.Add(currentNode.RightChild);
      newBucket.Add(currentNode.RightChild.Bucket);
      newFather.ReplaceChild(currentNode, currentLeaf);
      UpdateInOrderAdjacencyLinks();
      currentNode = currentNode.RightAdjacents[0];
   END WHILE
   UpdateAllLinks();
   Root.ComputeSizeInSubtree();
\end{verbatim}
\end{algorithm1}

The contraction (Alg. \ref{alg:contraction}) is carried out as follows: the last level of the PBT is deleted and every pair of adjacent buckets is merged into one bucket. In particular, each 2-level subtree $T_{2}$ at height $h = 1$ and its buckets $B_{1}$ and $B_{2}$ are replaced by a leaf $u$ and a bucket $B$, preserving, at the same time, the in-order adjacency. The leaf $u$ is basically the left leaf of $T_{2}$ and the new bucket $B$ contains the nodes of $B_{1}$, the root of $T_{2}$, the right leaf of $T_{2}$ and finally the nodes of bucket $B_{2}$. During the process, all in-order adjacency links and father-child links are updated. After the contraction has been carried out, the leaves reconstruct their routing tables and all binary nodes update their rightmost leaf.

It is obvious that these two operations are quite costly, since they involve a reconstruction of the overlay, but this reconstruction rarely happens.

\subsection{Single and Range Queries}
\label{subsec:search}

The search for an element $a$ may be initiated from any node $v$ at level $l$. If $v$ is a bucket node, then if its range contains $a$ the search terminates, otherwise the search is forwarded to the bucket representative, which is a binary node. In case $v$ is a node of the PBT (Alg. \ref{alg:search}), let $z$ be the node with range of values containing $a$, $a \in [x_z, x'_z]$ and assume w.l.o.g. that $x'_v < a$. The case where $x_v > a$ is completely symmetric. First, we perform a horizontal binary search at the level $l$ of $v$ using the routing tables, searching for a node $u$ with right sibling $w$ (if there is such sibling) such that $x'_u<a$ and $x_w>a$, unless $a$ is in the range of $u$ or in the range of any visited node of $l$ and the search terminates.

More specifically, we keep travelling to the right using the rightmost links of the routing tables of nodes in level $l$ (the most distant ones), until we find a node $q$ such that $x_q > a$, or until we have reached the rightmost node $qr$ of level $l$. If the first case is true, then $a$ is somewhere between $q$ and the last visited node in the left of $q$, so we start travelling to the left decreasing our travelling step by 1. We continue travelling left and right, gradually decreasing the travelling step, until we find the siblings $u$ and $w$ mentioned above in step = 0. If the second case is true, then $x'_{qr} < a$ and according to the in-order traversal the search is confined to the right subtree of $qr$.

\begin{algorithm1}
Search(integer element)
\label{alg:search}
\begin{verbatim}
   REQUIRE BinaryNode currentNode, step != 0
   BinaryNode leftMostNode, rightMostNode;
   /* horizontal search */
   WHILE step != 0 DO
      IF currentNode.Contains(element) THEN
         RETURN currentNode;
      END IF
      IF currentNode.UpperValue <= element THEN
         IF currentNode == RightMostNodeInLevel THEN
            RETURN currentNode.SearchSubtree(element);
         END IF
         leftMostNode = currentNode;
         IF rightMostNode != null THEN
            step--;
            currentNode = currentNode.RightAdjacents[step];
         ELSE
            currentNode = currentNode.RightAdjacents.Last;
            step = currentNode.RightAdjacents.Count;
         END IF
      ELSE IF currentNode.LowerValue > element THEN
         /*... similar to previous case ...*/
      END IF
   END WHILE
   /* step == 0, vertical search */
   BinaryNode leftSibling, rightSibling;
   IF currentNode.UpperValue <= element THEN
      leftSibling = currentNode;
      rightSibling = currentNode.RightAdjacents[0];
   ELSE IF currentNode.LowerValue > element THEN
      leftSibling = currentNode.LeftAdjacents[0];
      rightSibling = currentNode;
   END IF
   targetNode = leftSibling.SearchSubtree(element);
   IF targetNode == null THEN
      targetNode = leftSibling.SearchAncestor(element);
      IF targetNode == null THEN
         targetNode = rightSibling.SearchSubtree(element);
      END IF
   END IF
   RETURN targetNode;
\end{verbatim}
\end{algorithm1}

Having located nodes $u$ and $w$, the horizontal search procedure is terminated and the vertical search is initiated. Node $z$ will either be the common ancestor of $u$ and $w$ (Alg. \ref{alg:searchAncestor}), or in the right subtree rooted at $u$ (Alg. \ref{alg:searchSubtree}), or in the left subtree rooted at $w$. 
If $u$ is a binary node, it contacts the rightmost leaf $y$ of its subtree. If $x_y > a$ then an ordinary top down search from node $u$ will suffice to find $z$. Otherwise, node $z$ is in the bucket of $y$, or in its right in-order adjacent (this is also the common ancestor of $u$ and $w$), or in the subtree of $w$. If $z$ belongs to the subtree of $w$, a symmetric search is performed.

When $z$ is located, if $a$ is found in $z$ then the search was successful, otherwise $a$ is not stored in the structure. The search for an element $a$ is carried out in $O(\log{N})$ steps and is verified through experiments presented in this work.

A range query $[a, b]$ initiated at node $v$, invokes a search operation for element $a$. Node $z$ that contains $a$ returns to $v$ all elements in its range. If all elements of $u$ are reported then the range query is forwarded to the right adjacent node (in-order traversal) and continues until an element larger than $b$ is reached for the first time.

\begin{algorithm1}
SearchSubtree(integer element)
\label{alg:searchSubtree}
\begin{verbatim}
   REQUIRE BinaryNode currentNode
   IF currentNode.Contains(element) THEN
      RETURN currentNode;
   END IF
   IF currentNode IS Leaf THEN
      RETURN currentNode.Bucket.Search(element);
   END IF
   IF currentNode.UpperValue <= element THEN
      rightLeaf = currentNode.RightMostLeaf;
      IF rightLeaf.Contains(element) DO
         RETURN rightLeaf;
      END IF
      IF rightLeaf.UpperValue <= element THEN
         RETURN rightLeaf.Bucket.Search(element);
      ELSE
         /* search internal subtree */
         currentNode = currentNode.RightChild;
         WHILE currentNode != null DO
            IF currentNode.Contains(element) THEN
               RETURN currentNode;
            END IF
            IF currentNode.Node.UpperValue <= element THEN
               currentNode = currentNode.RightChild;
            ELSE IF currentNode.Node.LowerValue > element THEN
               currentNode = currentNode.LeftChild;
            END IF
         END WHILE
         RETURN currentNode.Bucket.Search(element);
      END IF
   ELSE IF currentNode.LowerValue > element THEN
      /*... similar to previous case ...*/
   END IF
   RETURN null;
\end{verbatim}
\end{algorithm1}

\begin{algorithm1}
SearchAncestor(integer element)
\label{alg:searchAncestor}
\begin{verbatim}
   REQUIRE BinaryNode currentNode
   IF currentNode.UpperValue <= element THEN
      IF currentNode IS Leaf THEN
         ancestor = currentNode.RightInOrderAdjacent;
      ELSE
         ancestor = currentNode.RightMostLeaf.RightInOrdetAdjacent;
      END IF
   ELSE IF currentNode.LowerValue > element THEN
      /*... similar to previous case ...*/
   END IF
   IF ancestor.Contains(element) THEN
      RETURN ancestor;
   END IF
   RETURN null;
\end{verbatim}
\end{algorithm1}

\subsection{Element Insertions and Deletions}
\label{subsec:elements}
\subsubsection{Handling element updates.}
Assume that an update operation (insertion/ deletion) is initiated at node $v$ involving element $a$. By invoking a search operation, node $u$ with range containing element $a$ is located and the update operation is performed on $u$.
Element $a$ is inserted in $u$ or is deleted from $u$, depending on the request.

In order to apply the weight-based mechanism for load balancing, the element should be inserted in a bucket node (similar to node joins) or in a leaf. 
However, node $u$ can be any node in the structure, even an internal node of PBT. 
In case $u$ is a bucket node or a leaf, $a$ is inserted to $u$ and no further action is necessary.
If $u$ is an internal node of the PBT, element $a$ is inserted in $u$ and then the first element of $u$ (note that elements into nodes are sorted) is removed from $u$ and inserted to $q$, the last node of the bucket of the left adjacent of $u$, in order to preserve the sequence of elements in the in-order traversal. This way, the insertion has been shifted to a bucket node. The case of element deletion is similar.

After an element update in leaf $u$ or in its bucket, the weight-based mechanism is activated and updates the weights by $\pm 1$ on the path from leaf $u$ to the root, as long as Invariants \ref{inv:1} or \ref{inv:2} do not hold. When the first node $w$ is accessed for which Invariants \ref{inv:1} and \ref{inv:2} hold, then the nodes in the subtree of its child $q$ in the path, have their weights recomputed. Afterwards, the mechanism traverses the path from leaf $u$ to the root, in order to find the first node (if such a node exists) for which Invariant \ref{inv:3} is violated and performs a load-balancing in its subtree.

\subsubsection{Load Balancing.}\label{subsec:loadbalancing}
The load-balancing mechanism guarantees that if there are $w(v)$ elements in total in the subtree of $v$ of size $|v|$ (total number of nodes in the subtree of $v$ including $v$), then after load-balancing each node stores either $\left\lfloor \frac{w(v)}{|v|} \right\rfloor$ or $\left\lfloor \frac{w(v)}{|v|} \right\rfloor+1$ elements. The load-balancing cost is $O(\log{N})$, which is verified through experiments, presented analytically in this work.

The load-balancing in the subtree of $v$ is carried out as follows (Alg. \ref{alg:loadbalancing}). We assume that $v$ has $|v|$ nodes in its subtree. A bottom-up computation of the weights in all nodes of the subtree is performed, in order to determine the weight $w(v)$ of $v$.  Then, the first $k$ nodes will contain $\left\lfloor \frac{w(v)}{|v|} \right\rfloor+1$ elements after load-balancing, where $k = w(v)\bmod{|v|}$. The remaining $|v| - k$ nodes will contain $\left\lfloor \frac{w(v)}{|v|} \right\rfloor$ elements. The redistribution starts from the rightmost node $w$ of the rightmost bucket $b$ and it is performed in an in-order fashion.

We assume that $w$ has $m$ extra elements which must be transferred to other nodes. Node $w$ has a link to node $w'$ in which the $m$ extra elements should be inserted. In order to locate node $w'$, we take into consideration the following cases: (i) if $w$ is a bucket node, $w'$ is its left node, unless $w$ is the first node of the bucket and then $w'$ is the bucket representative, (ii) if $w$ is a leaf, then $w'$ is the left in-order adjacent of $w$ and (iii) if $w$ is an internal binary node, then its left in-order adjacent is a leaf and $w'$ is the last node of its bucket. Having located $w'$, the first $m$ extra elements of $w$ are removed from $w$ and are added to the end of the element queue of $w'$, in order to preserve the indexing structure of the tree. Then, the ranges of $w$ and $w'$ are updated respectively.

The case where $m$ elements must be transferred to node $w$ from node $w'$ is completely symmetric. In general, the last $m$ elements of $w'$ are removed from $w'$ and are inserted in the first $m$ places in the element queue of $w$. The intriguing part is when $w'$ contains less elements than the $m$ elements that $w$ needs. In this case, \textit{dest} travels towards the leftmost node of the subtree, following the in-order traversal, until $m \leq \sum_{i=1}^{s}{e(u_{i})}$, where $e(u_i)$ is the number of elements of the $i-th$ node on the left. Then, elements of node $u_s$ are transferred to $u_{s-1}$, elements from $u_{s-1}$ are transferred to $u_{s-2}$ and so on, until \textit{dest} goes backwards to $w'$ and $m$ elements are moved from $w'$ into $w$.

\begin{algorithm1}
LoadBalanceSubtree()
\label{alg:loadbalancing}
\begin{verbatim}
   REQUIRE BinaryNode node
   node.ComputeSizeInSubtree();
   node.ComputeWeightInSubtree();
   newNodeDensity =  node.Weight / node.Size;
   leftMostLeaf = node.LeftMostLeaf;
   currentNode = node.RightMostLeaf;
   WHILE currentNode != leftMostLeaf DO
      destinationNode = node.FindDestinationNode();
      IF currentNode.Elements > newNodeDensity THEN
         elementsToMove = currentNode.Elements - newNodeDensity;
         SplitElements(currentNode, destinationNode, elementsToMove);
      ELSE IF currentNode.Elements < newNodeDensity THEN
         elementsToMove = newNodeDensity - currentNode.Elements;
         tempDestNode = destinationNode;
         WHILE tempDestNode != leftMostLeaf AND availableElements < 
         elementsToMove DO
            availableElements += tempDestNode.Elements;
            tempDestNode = tempDestNode.FindDestinationNode();
         END WHILE
         tempElementsToMove = tempDestNode.Elements - 
         (totalElementsAvailable - elementsToMove);
         tempSourceNode = tempDestNode.FindSourceNode();
         WHILE tempDestNode != currentNode DO
            SplitElements(tempDestNode, tempSourceNode, tempElementsToMove);
            tempDestNode = tempSourceNode;
            tempElementsToMove = tempDestNode.Elements;
            tempSourceNode = tempSourceNode.FindSourceNode();
         END WHILE
      END IF
      currentNode = destinationNode;
   END WHILE
\end{verbatim}
\end{algorithm1}

\subsection{Fault Tolerance}
\label{subsec:fault}

Searches and updates in the $D^3$-Tree do not tend to favour any node, and in particular nodes near the root, which are therefore not crucial and their failure will not cause more problems than the failure of any node. 
However, a single node can be easily disconnected from the overlay, when all nodes with which it is connected fail. This means that 4 failures (two adjacent nodes and two children) are enough to disconnect the root.
The most easily disconnected nodes are those which are near the root, since their routing tables are small in size.

When a node $w$ discovers that $v$ is unreachable, the network initiates a node withdrawal procedure by reconstructing the routing tables of $v$, in order $v$ to be removed smoothly, as if $v$ was departing. If $v$ belongs to a bucket, it is removed from the structure and the links of its adjacent nodes are updated.

In case $v$ is an internal binary node, its right adjacent node $u$ is first located, making use of Lemma\ref{lem:properties}, in order to replace $v$.
More specifically, we assume that node $w$ discovered that $v$ is unreachable during some operation.
Taking into account all possible relative positions between $w$ and $v$, we have the following cases, in which we want to locate the right child $q$ of $v$ that will lead as to $u$. First, if $w$ and $v$ are on the same level, by Lemma\ref{lem:properties}(i) we locate $q$ and thus the right adjacent node $u$ of $v$ is the leftmost leaf in the subtree of $q$. Being more clear, if $w$ is connected to $v$ by the $i$-th link of its routing table, then its right child is connected to $q$ by the $(i+1)$-th link of its routing table. 
Second, if $w$ is the father of $v$, by Lemma\ref{lem:properties}(i) its left (right) child $p$ has a link to the missing node $v$ and the right child of $p$ has a link to $q$, so $u$ is located. Third, if $w$ is the left (right) child of $v$, then $u$ is easily located.

In case $v$ is a leaf, then it should be replaced by the first node $u$ in its bucket. However, in D$^3$-Tree predecessor, if a leaf was found unreachable, contacting its bucket would be infeasible, since the only link between $v$ and its bucket would have been lost. This weakness was eliminated in D$^3$-Tree, by maintaining multiple links towards each bucket, distributed in exponential steps (in the same way as the horizontal adjacency links). 
This way, when $w$ is unable to contact $v$, it contacts directly the first node of its bucket $u$ and $u$ replaces $v$.

In any case, the elements stored in $v$ are lost. Moreover, the navigation data of $u$ (left adjacent of $v$) are copied to the first node $z$ in its bucket which takes its place, and $u$ has its routing tables recomputed.

\subsection{Single Queries with Node Failures}
\label{subsec:queriesFailures}

The problem of searching an element in a network where a number of nodes have fallen, introduces some very intriguing aspects and it can be considered as two-dimensional, since the search must be both successful and cost effective.
A successful search for element $a$ refers to locating the target node $z$ for which $a \in [x_z, x'_z]$. An unsuccessful search refers to the cases where (i) $z$ is unreachable and, (ii) there is a path to $z$ but the search algorithm couldn't follow it to locate $z$, due to failures of intermediate nodes. The D$^3$-Tree predecessor doesn't provide a search algorithm in case of node failures, since it doesn't sufficiently confront the structure's fault tolerance. In the following, we present the key features of our search algorithm, mostly through examples, due to the complexity of its implementation.

The search procedure is similar to the simple search described in section \ref{subsec:search}. One difference in horizontal search lies in the fact that if the most distant right adjacent of $v$ is unreachable, $v$ keeps contacting its right adjacent nodes by decreasing the step by 1, until it finds node $q$ which is reachable.
If all right adjacents are unreachable, $v$ contacts its left adjacents, afterwards it tries to contact its children, its father and as a last chance, when all other nodes have failed, it contacts its left/right in-order adjacents and its left/rightmost leaf. Contacting children, in-order adjacents and leaves means a change in the search level.

\begin{figure}
\centering
\includegraphics[width=2.8in]{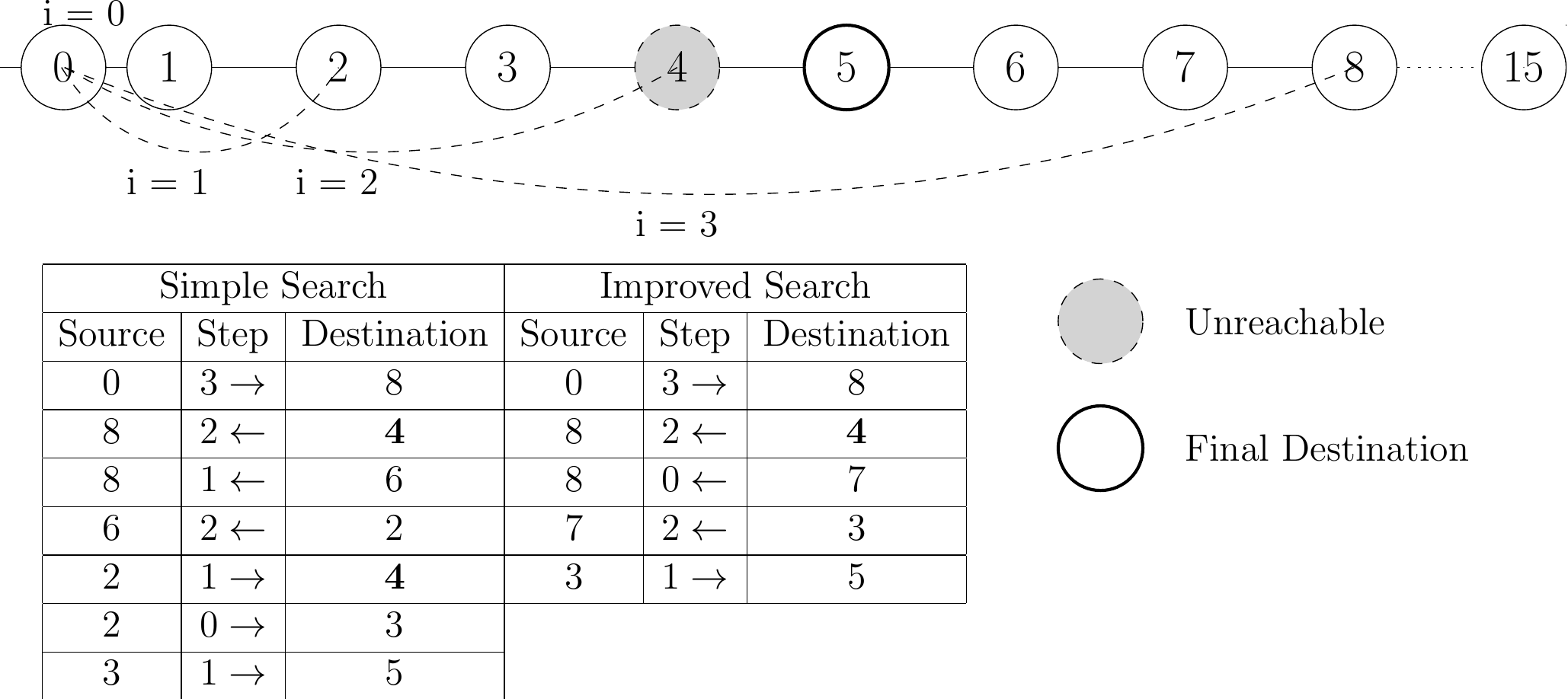}
\caption{Example of binary horizontal search with node failures}
\label{fig:graph01}
\end{figure}

In case $x'_q < a$ the search continues to the right using the most distant right adjacent of $q$, otherwise the search continues to the left and $q$ contacts its most distant left adjacent $p$ which is in the right of $v$. If $p$ is unreachable, $q$ doesn't decrease the travelling step by 1, but contacts directly its nearest left adjacent (at step = 0) and asks it to search to the left. This improvement reduces the number of messages that are meant to fail, because of the exponential positions of nodes in routing tables and the nature of binary horizontal search. For example, in Fig. \ref{fig:graph01}, search starts from $v_0$ and $v_8$ contacts $v_7$ since $v_4$ has fallen. No node contacts $v_4$ from then on and the number of messages is reduced by 2.

\begin{figure}
\centering
\includegraphics[width=2.5in]{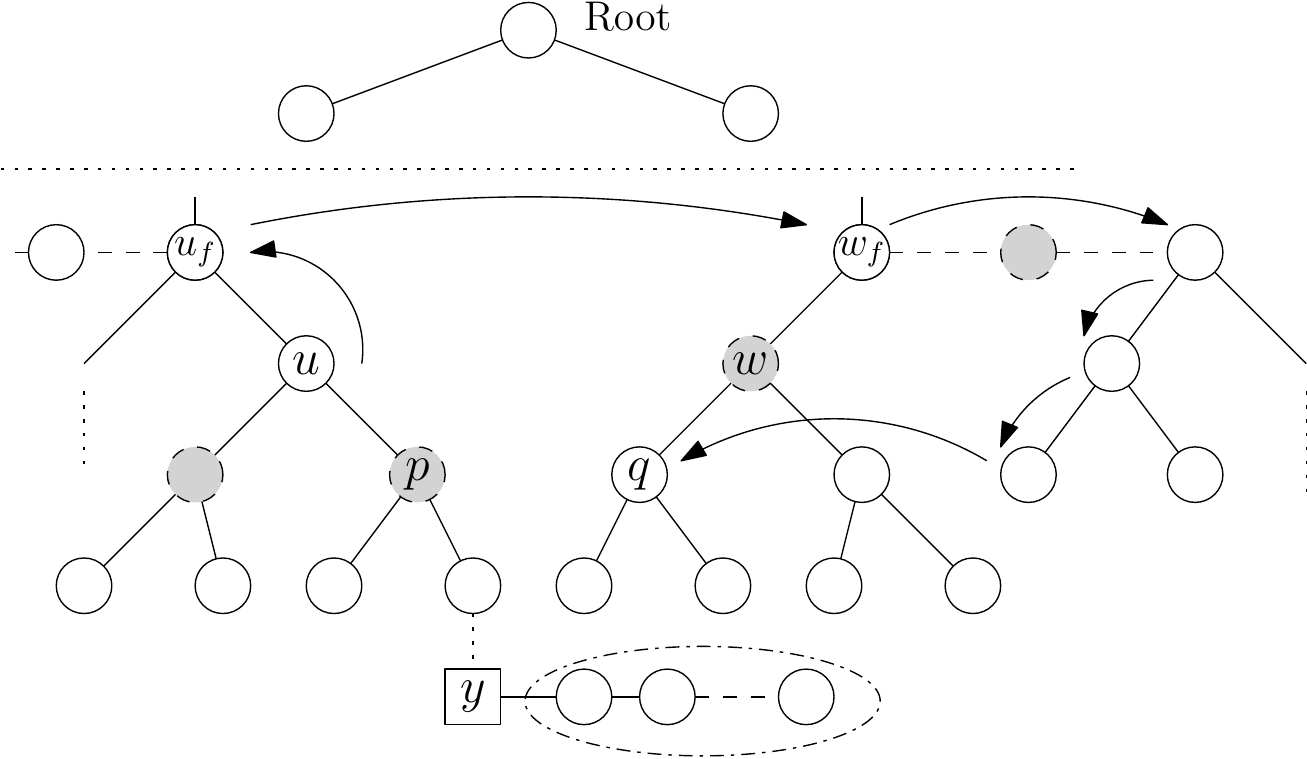}
\caption{Example of vertical search between $u$ and unreachable $w$}
\label{fig:tree01}
\end{figure}

A vertical search to locate $z$ is always initiated between two siblings $u$ and $w$, which are either both active, or one of them is unreachable, as shown in Fig \ref{fig:tree01} where the left sibling $u$ is active and $w$, the right one, is unreachable. In both cases, first we search into the subtree of the active sibling, then we contact the common ancestor and then, if the other sibling is unreachable, the active sibling tries to contact its corresponding child (right child for left sibling and left child for right sibling). When the child is found the search is forwarded to its subtree.
We assume (w.l.o.g.) that the left sibling $u$ is active and $w$, the right one, is unreachable, as shown in Fig. \ref{fig:tree01}. First $u$ contacts its rightmost leaf $y$ of its subtree. If $y$ is reachable and $x_y > a$ or if $y$ is unreachable, then an ordinary top down search from node $u$ is initiated to find $z$. If a node in the searching path is unreachable, a mechanism, which is described below, is activated to contact its children. Otherwise, if $x'_y < a$, node $z$ is in the bucket of $y$, or in its right in-order adjacent (this is also the common ancestor of $u$ and $w$), or in the left subtree of $w$.

In general, when node $u$ wants to contact the left (right) child of unreachable node $w$, the contact is accomplished through the routing table of its own left (right) child. If its child is unreachable, (Fig. \ref{fig:tree01}), then $u$ contacts its father $u_f$ and $u_f$ contacts the father of $w$, $w_f$, using Lemma\ref{lem:properties}(i), unless $u$ and $w$ have the same father.
Then $w_f$, using Lemma\ref{lem:properties}(ii) twice in succession, contacts its grandchild through its left and right adjacents and their grandchildren.

In case initial node $v$ is a bucket node, then if its range contains $a$ the search terminates, otherwise the search is forwarded to the bucket representative. If the bucket representative has fallen, the bucket contacts its other representatives right or left, until it finds a representative that is reachable. The procedure continues as described above for the case of a binary node.


\section{The ART$^+$ structure}
\label{sec:art+}

In this section we briefly describe and present the theoretical background of ART$^+$. ART$^+$ is similar to its predecessor, ART\cite{stpstm2012:art} regarding the structure's outer level. Their difference, which introduces performance enhancements, lies in the fact that each cluster-peer of ART$^+$ is structured as a D$^3$-Tree\cite{bstz14}.

\subsubsection{Building the ART$^+$ structure.}
\begin{figure}
\centering
\includegraphics[width=4in]{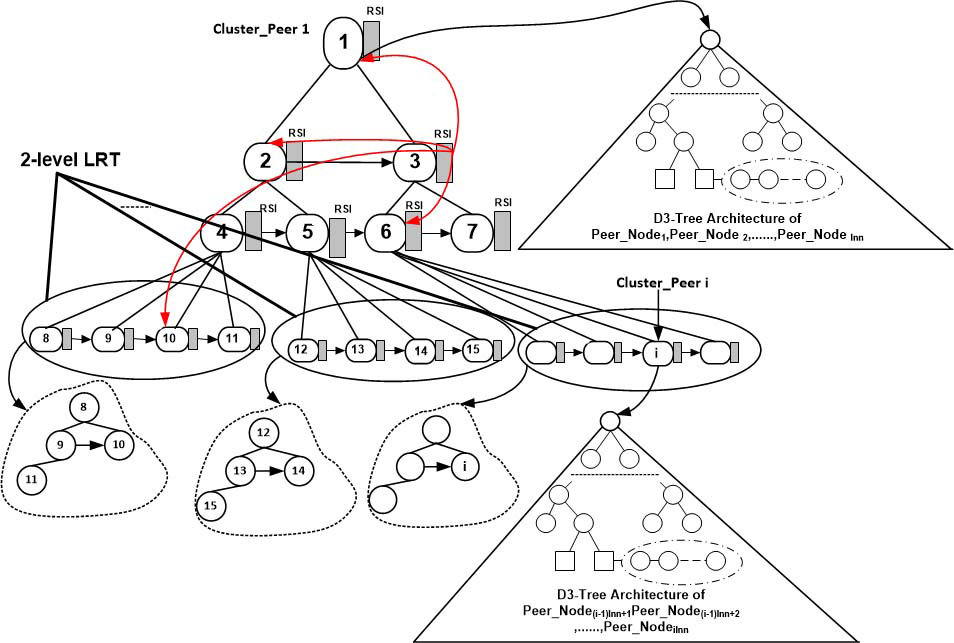}
\caption{The ART$^+$ structure for b = 2.}
\label{fig:art}
\end{figure}

The backbone structure of ART$^+$ is similar to LRT\footnote{LRT: Level Range Tree}, in which some interventions have been made to improve its performance and increase the robustness of the whole system. ART$^+$ is built by grouping cluster-peers having the same ancestor and organizing them in a tree structure recursively. A cluster-peer is defined as a bucket of ordered peers. The innermost level of nesting (recursion) will be characterized by having a tree in which no more than $b$ cluster-peers share the same direct ancestor, where $b$ is a double-exponentially power of two (e.g. 2, 4, 16,...). Thus, multiple independent trees are imposed on the collection of cluster-peers. The height of ART$^+$ is $O(\log{\log_b{N}})$ in the worst case. The ART$^+$ structure remains unchanged w.h.p. Figure \ref{fig:art} illustrates a simple example, where $b = 2$.

The degree of the cluster-peers at level $i > 0$ is $d(i) = t(i)$, where $t(i)$ indicates the number of cluster-peers at level $i$. It holds that $d(0) = b$ and $t(0) = 1$. At initialization step, the 1st peer, the $(\ln{n} + 1)-th$ peer, the $(2 \cdot \ln{n} + 1)-th$ peer and so on are chosen as bucket representatives, according to the balls in bins combinatorial game presented in \cite{kmsttz2003}.
Let $n$ be $w$-bit keys, $N$ be the total number of peers and $N'$ be the total number of cluster-peers. Each peer with label $i$ (where $1 \leq i \leq N$) of a random cluster, stores ordered keys that belong in the range $[(i - 1) \ln{n}, i \ln{n} - 1]$, where $N = n / \ln{n}$. Each cluster-peer with label $i'$ (where $1 \leq i' \leq N'$) stores ordered peers with sorted keys belonging in the range $[(i'- 1) \ln^2{n}, . . . , i' \ln^2{n} - 1]$, where $N' = n / \ln^2{n}$ or $N' = N / \ln{n}$ is the number of cluster-peers.
 
ART$^+$ stores cluster-peers only, each of which is structured as an independent decentralized architecture, which changes dynamically after node join/leave and element insert/delete operations inside it. In contract to its predecessor, ART, whose inner level was structured as a BATON$^*$, each cluster-peer of ART$^+$ is structured as a D$^3$-Tree.
Each cluster-peer is equipped with a routing table named Random Spine Index (RSI), which stores pointers to cluster-peers belonging to a random spine of the tree (instead of the LSI\footnote{LSI: Left Spine Index} of LRT which stores pointers to the peers of the left-most spine). Moreover, instead of using fat CI\footnote{CI: Collection Index} tables, which store pointers to the collections of peers presented at the same level, the appropriate collection of cluster-peers is accessed by using a 2-level LRT structure. In ART$^+$, the overlay of cluster-peers remains unaffected in the expected case w.h.p. when peers join or leave the network.

\subsubsection{Load Balancing.}
The operation of join/leave of peers inside a cluster-peer is modelled as the combinatorial game of bins and balls presented in \cite{kmsttz2003}. In this way, for an $\mu(\cdot)$ random sequence of join/leave peer operations, the load of each cluster peer never exceeds $\Theta(\log{N})$ size and never becomes zero in expected w.h.p. case. In skew sequences, though, the load of each cluster-peer may become $\Theta(N)$ in worst case.
The load-balancing mechanism for a D$^3$-tree structure, as described previously, has an amortized cost of $O(\log{K})$, where $K$ is the total number of nodes in the D$^2$-tree. Thus, in an ART$^+$ structure, the cost of load-balancing is $O(\log{\log{N}})$ amortized.

\subsubsection{Routing Overhead.}
The 2-level LRT is an LRT structure over $\log^{2c}{Z}$ buckets each of which organizes $Z / \log^{2c}{Z}$ collections in a LRT manner, where $Z$ is the number of collections at current level and $c$ is a big positive constant. As a consequence, the routing information overhead becomes $O(N^{1/4} / \log^c{N})$ in the worst case.

\subsubsection{Search Algorithms.}
Since the structure's maximum number of nesting levels is $O(\log_b{\log{N}})$ and at each nesting level $i$ we have to apply the standard LRT structure in $N^{1 / {2^i}}$ collections, the whole searching process requires $O(\log_b^2{\log{N}})$ hops. Then, we have to locate the target peer by searching the respective decentralized structure. Through the polylogarithmic load of each cluster peer, the total query complexity $O(\log_b^2{\log{N}})$ follows. Exploiting now the order of keys on each peer, range queries require $O(\log_b^2{\log{N}} +|A|)$ hops, where $|A|$ is the answer size.

\subsubsection{Join/Leave Operations.}
A peer $u$ can make a join/leave request to a peer $v$, which is located at cluster peer $W$. Since the size of $W$ is bounded by a $polylogN$ size in expected w.h.p. case, the peer join/leave can be carried out in $O(\log{\log{N}})$ hops.
 The outer structure of ART$^+$ remains unchanged w.h.p. as mentioned before, but each D$^3$-tree structure changes dynamically after peer join/leave operations. According to D$^3$-Tree performance evaluation, the peer join/leave can be carried out in $O(\log{\log{N}})$ hops.

\subsubsection{Node Failures and Network Restructuring.}
In the ART$^+$ structure, similarly to ART, the overlay of cluster-peers remains unchanged in the expected case w.h.p., so in each cluster-peer the algorithms for node failure and network restructuring are according to the decentralized architecture used. D$^3$-Tree is a highly fault-tolerant structure, since it supports procedures for node withdrawal and handles massive node failures efficiently.


\section{Experimental Study}
\label{sec:experiments}

\begin{figure}
\centering
\includegraphics[width=4.8in]{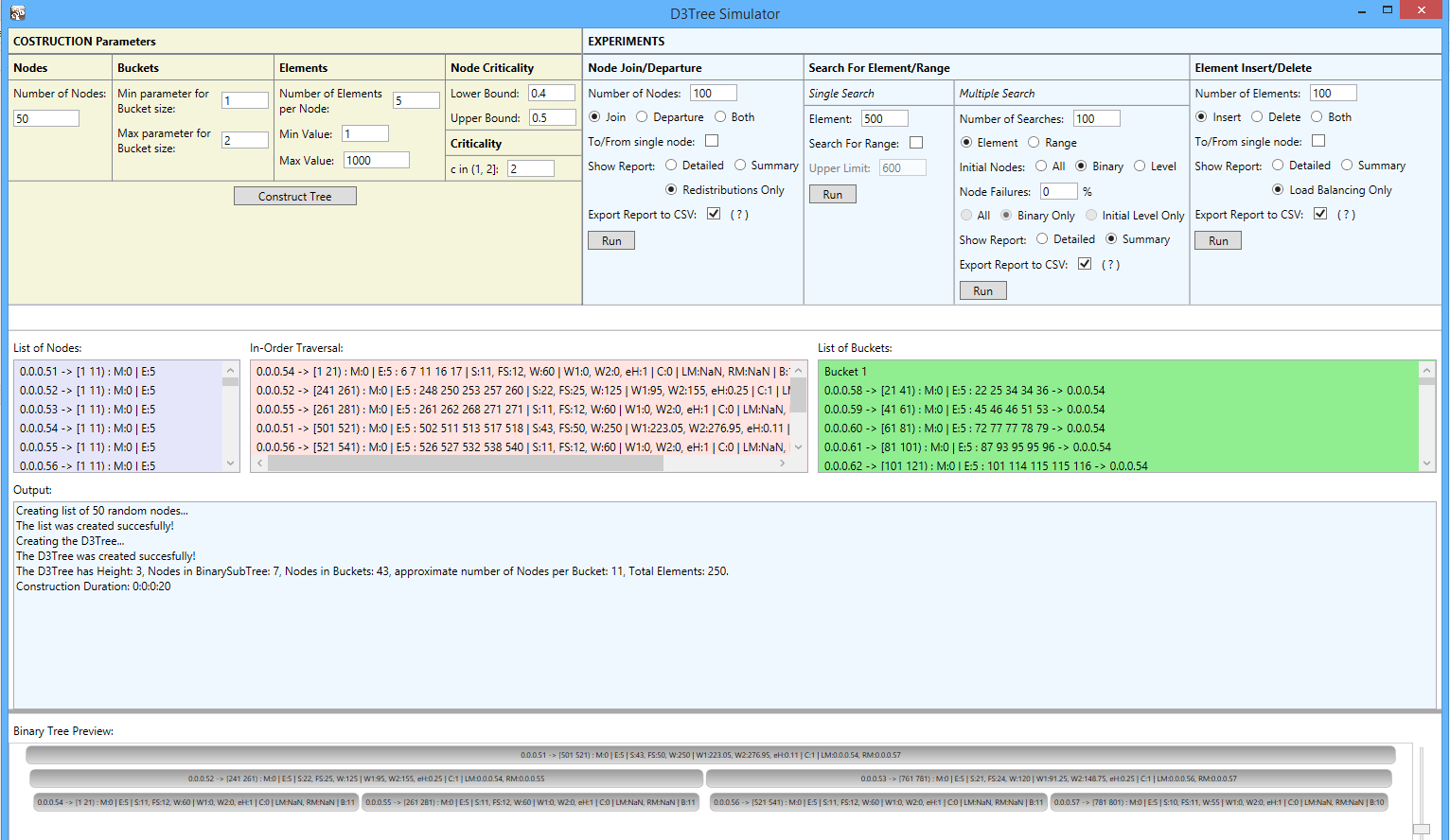}
\caption{The D$^3$-Tree simulator user interface.}
\label{fig:ss}
\end{figure}

We built a simulator\footnote{Our simulator is a standalone desktop application, developed in Visual Studio 2010, available in \url{https://github.com/sourlaef/d3-tree-sim}} with a user friendly interface and a graphical representation of the structure, to evaluate the performance of D$^3$-Tree.
In Fig. \ref{fig:ss} we present the user interface of the D$^3$-Tree simulator. At the top left, the user can construct a new tree structure after setting the tree parameters. At the top right, the user can conduct experiments regarding node joins/departures, element insertions/deletions, single and range queries, setting the parameters for each experiment. In the centre of the screen, useful information is displayed regarding the tree construction and experiments. At the bottom, a graphical representation of the structure is displayed.

\subsection{Performance Evaluation for D$^3$-Tree}

To evaluate the cost of operations, we ran experiments with different number of nodes $N$ from 1,000 to 10,000, in order to be directly compared to BATON, BATON$^*$ and P-Ring.
BATON$^*$ is a state-of-the-art decentralized architecture and P-Ring outperforms DHT-based structures in range queries and achieves a slightly better load-balancing performance compared to Baton$^*$.
For a structure of $N$ nodes, 1000 x $N$ elements where inserted. We used the number of passing messages to measure the performance of the system.

\subsubsection{Cost of Node Joins/Departures.}
\begin{figure}[]
\centering
    \subfloat[average case\label{subfig-1:exp_nodes_ext}]{%
      \includegraphics[width=2.2in]{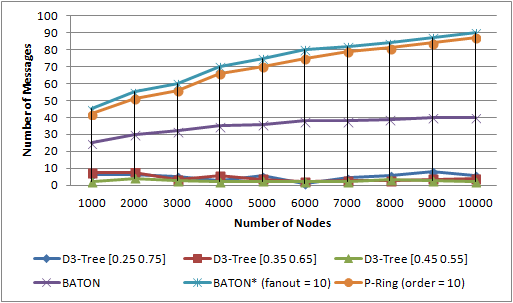}
    }
    \hfill
    \subfloat[worst case\label{subfig-2:exp_nodes_ext}]{%
      \includegraphics[width=2.4in]{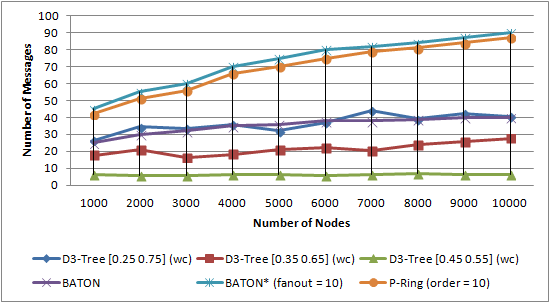}
    }
    \\
    \subfloat[redistribution rate of winner D$^3$-Tree\label{subfig-3:exp_nodes_ext}]{%
      \includegraphics[width=2.3in]{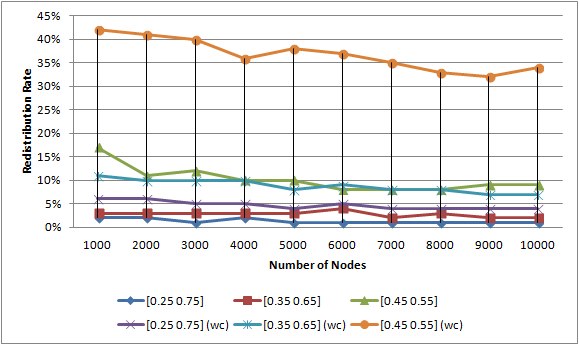}
    }
    \hfill
    \subfloat[redistr/ion height of winner D$^3$-Tree\label{subfig-4:exp_nodes_ext}]{%
      \includegraphics[width=2.3in]{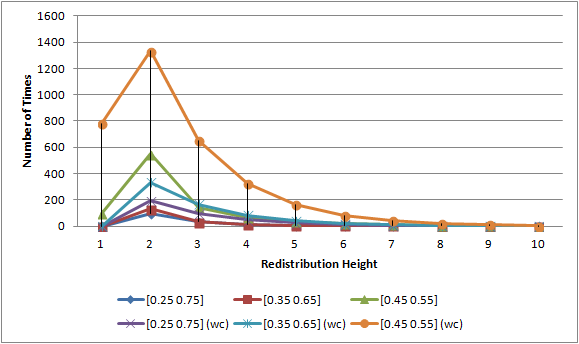}
    }
    \caption{Node Update operations}
    \label{fig:exp_nodes_ext}
  \end{figure}

To measure the network performance for node updates, we conducted experiments for three different value ranges of \textit{node criticality}: [0.25 0.75], [0.35 0.65] [0.45 0.55]. For a network of $N$ initial nodes, we performed $2N$ node updates. Figure \ref{subfig-1:exp_nodes_ext} shows the average amortized redistribution cost, while \ref{subfig-2:exp_nodes_ext} depicts the same metrics in worst case, where the same node (the leftmost leaf) is the host node for all node joins.
Note that we have taken into account only the amortized cost of node joins causing redistributions, since otherwise, the amortized cost is negligible.
Figure \ref{subfig-3:exp_nodes_ext} shows the redistribution rate for D$^3$-Tree in average and worst case.

Through experiments, we observed that even in the worst case scenario, the D$^3$-Tree node update and redistribution mechanism achieves a better amortized redistribution cost, compared to that of BATON, BATON$^*$ and P-Ring.
We also observed that in the average case, during joins and departures of nodes, the Invariant \ref{inv:nodecriticality} is rarely violated to invoke a redistribution operation (Fig. \ref{subfig-3:exp_nodes_ext}). This also depends on the range in which node criticality belongs. When the range is narrowed, more redistributions take place during node updates, but the amortized cost is low, since the majority of redistributions occur in subtrees of low height, as shown in Fig. \ref{subfig-4:exp_nodes_ext}. This is more obvious in worst case. There, when we use wide ranges, more node joins take place before a redistribution occurs, making the redistribution operation more costly, since a great number of nodes have been cumulated into the bucket of the leftmost leaf.

Figure \ref{subfig-4:exp_nodes_ext} shows in detail the allocation of redistribution height for different node criticality ranges in a network of 10,000 initial nodes. We observe that in worst case the number of redistributions is more than twice the number of redistributions of the average case.

\subsubsection{Cost of Element Insertions/Deletions.}

\begin{figure}[]
\centering
    \subfloat[average messages\label{subfig-1:exp_elements_ext}]{%
      \includegraphics[width=2.3in]{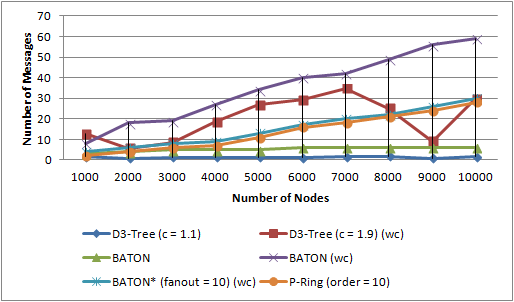}
    }
	\hfill 
    \subfloat[load-balancing rate of winner D$^3$-Tree\label{subfig-2:exp_elements_ext}]{%
      \includegraphics[width=2.3in]{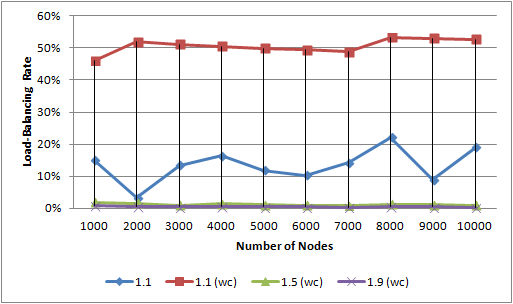}
    }
    \\ 
    \subfloat[load-balancing height of winner D$^3$-Tree\label{subfig-3:exp_elements_ext}]{%
      \includegraphics[width=2.4in]{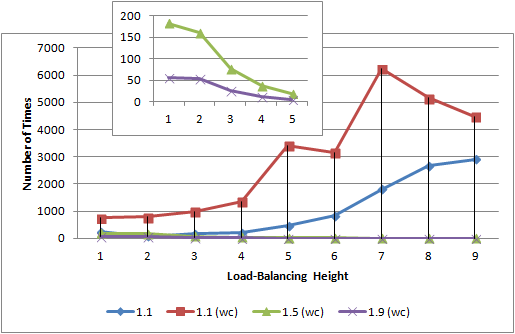}
    }
    \caption{Element Update operations}
    \label{fig:exp_elements_ext}
  \end{figure}

To measure the network performance for the operation of element updates, we conducted experiments for three different value ranges of \textit{criticality}: [0.90 1.10], [0.67 1.50], [0.53 1.90] formed from values of constant $c = \{1.1, 1.5, 1.9\}$ correspondingly. For a network of $N$ nodes and $n$ elements, we performed $n$ element updates. Figure \ref{subfig-1:exp_elements_ext} shows the average amortized load-balancing cost, while Fig. \ref{subfig-2:exp_elements_ext} shows the load-balancing rate. Both average cases and worst cases are depicted in the same graph. The average cases for $c$ values of 1.5 and 1.9 led to negligible amortized cost so they were disregarded. In worst case, the same node (the leftmost leaf) is the host node for all element insertions. Note that we have taken into account only the amortized cost of element insertions causing load-balancing operations, since otherwise, the amortized cost is negligible.

Conducting experiments, we observed that in the average case, the D$^3$-Tree outperforms BATON, BATON$^*$ and P-Ring. However, in D$^3$-Tree's worst case, the load-balancing performance is degraded compared to BATON$^*$ of \textit{fanout} = 10 and P-Ring.
Moreover, we observed that in the average case, during element updates, the Invariant \ref{inv:3} is rarely violated to invoke the load-balancing mechanism (Fig. \ref{subfig-2:exp_elements_ext}). This also depends on the value range of criticality. When the range is narrowed, more load-balancing operations take place during element updates, but the amortized cost is low since the subtree isn't very imbalanced, although the majority of load-balancing operations occur in subtrees of high height, as shown in Figure \ref{subfig-3:exp_elements_ext}. On the other hand, when we use wide ranges, more element updates take place before the load-balancing mechanism is activated, leading to more frequent and costly operations of load-balancing.

\subsubsection{Cost of Element Search with/without Node Failures.}

\begin{figure}[]
\centering
    \subfloat[average messages without failures\label{subfig-1:exp_sq}]{%
      \includegraphics[width=2.3in]{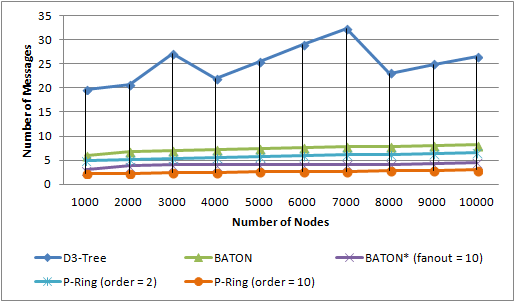}
    }
	\hfill
	\subfloat[effect of massive failure\label{subfig-2:exp_sq}]{%
      \includegraphics[width=2.3in]{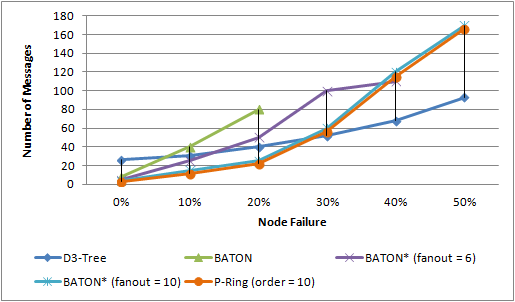}
    }
    \\
    \subfloat[average messages of D$^3$-Tree\label{subfig-3:exp_sq}]{%
      \includegraphics[width=2.3in]{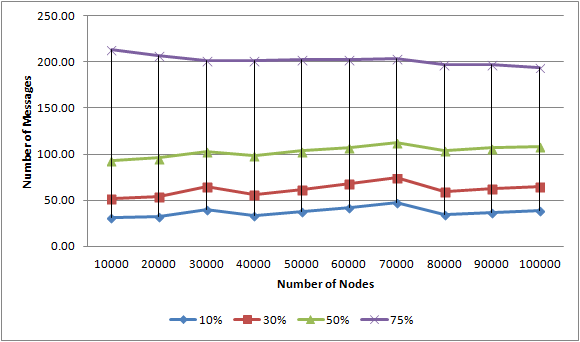}
    }
	\hfill
    \subfloat[success percentage of D$^3$-Tree\label{subfig-4:exp_sq}]{%
      \includegraphics[width=2.3in]{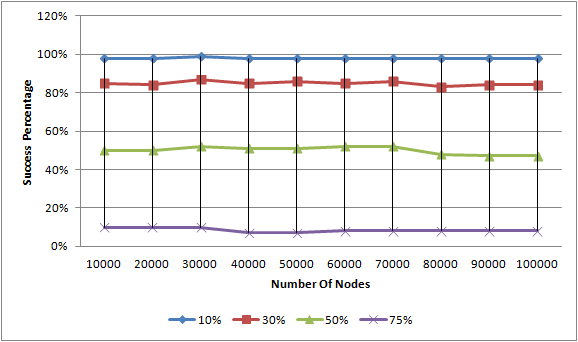}
    }
    \caption{Single Queries without/with node failures}
    \label{fig:exp_sq}
  \end{figure}

To measure the network performance for the operation of single queries, we conducted experiments in which for each $N$, we performed $2M$ ($M$ is the number of binary nodes) searches. The search cost is depicted in Fig. \ref{subfig-1:exp_sq}. An interesting observation here was that although the cost of search in D$^3$-Tree doesn't exceed $2 \cdot \log{N}$, it is higher that the cost of BATON, BATON$^*$ and P-Ring. This is due to the fact that when the target node is a \textit{Bucket} node, the search algorithm, after locating the correct leaf, performs a serial search into its bucket to locate it.

To measure the network performance for the operation of element search with node failures, we conducted experiments for different percentages of node failures: 10$\%$, 20$\%$, 30$\%$, 50$\%$ and 75$\%$. For each $N$, we performed $2M$ ($M$ is the number of binary nodes) searches divided in 4 sets. A different set of nodes failed in each of the 4 sets.
Figure \ref{subfig-2:exp_sq} depicts the increase in search cost when massive failures of nodes take place in D$^3$-Tree, BATON, different fanouts of BATON$^*$ and P-Ring. We observe that D$^3$-Tree maintains low search cost, compared to the other structures, even for a failure percentage $ \geq 30\%$.

Details about the behaviour of the enhanced search mechanism of D$^3$-Tree in case of node failures, are depicted in Fig. \ref{subfig-3:exp_sq} and Fig. \ref{subfig-4:exp_sq}, which show the average number of messages required and the success percentage respectively.
Experimenting, we observed that when the node failure percentage is small (10$\%$ to 15$\%$), the majority of single queries that fail are the ones whose elements belong to failed nodes. When the number of failed nodes increases, single queries are not always successful, since the search mechanism fails to find a path to the target node although the node is reachable. However, even for the significant node failure percentage of $30\%$, our search algorithm is $85\%$ successful, confirming thus our claim that the proposed structure is highly fault-tolerant.

\subsection{Performance Evaluation for ART$^+$}

We also evaluated the performance of ART$^+$ structure and compared it to its predecessor, ART \cite{stpstm2012:art}. Each cluster peer of the ART and ART$^+$ is a BATON$^*$ and D$^3$-Tree structure respectively. BATON$^*$ was implemented and evaluated in \cite{jotvz06}, while ART was evaluated in \cite{stpstm2012:art}, using the Distributed Java D-P2P-Sim simulator presented in \cite{spstm:p2psim}. The source code of the whole evaluation process, which showcases the improved performance, scalability, and robustness of ART over BATON$^*$ is publicly available\footnote{\url{http://code.google.com/p/d-p2p-sim/}}. For the performance evaluation of ART$^+$, we used the D$^3$-Tree simulator.

To evaluate the performance of ART and ART$^+$ for the search and load-balancing operations, we ran experiments with different number of total nodes $N$ from 50,000 to 500,000. As proved in \cite{stpstm2012:art}, each cluster peer stores no more than $0.75\log^2{N}$ peers in smooth distributions (normal, beta, uniform) and no more than $2.5\log^2{N}$ peers in non-smooth distributions (powlow, zipfian, weibull). Moreover, we inserted elements equal to the network size multiplied by 2000, which are numbers from the universe $[1...1,000,000,000]$.  We used the number of passing messages to measure the performance.

Note here that, as proved in \cite{stpstm2012:art}, ART outperforms BATON$^*$ in search operations, except for the case where $b=2$. Moreover, ART achieves better load-balancing compared to BATON$^*$, since the cluster-peer overlay remains unaffected w.h.p. through joins/departures of peers and the load-balancing performance is restricted inside a cluster-peer. Consequently, in this work, ART$^+$ is compared directly to ART.

\subsubsection{Cost of Search Operations.}

\begin{figure}[]
\centering
    \subfloat[case b=2\label{subfig-1:exp_search}]{%
      \includegraphics[width=2.3in]{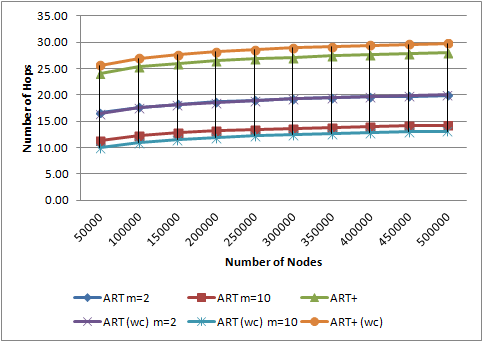}
    }
	\hfill
    \subfloat[case b=4\label{subfig-2:exp_search}]{%
      \includegraphics[width=2.3in]{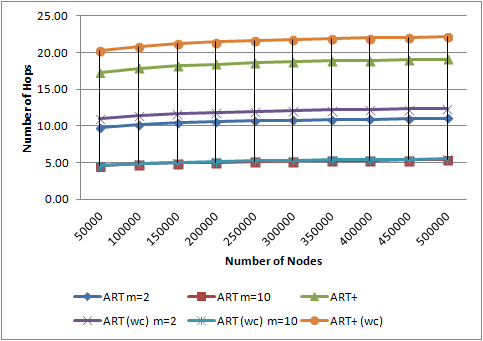}
    }
	\\
    \subfloat[case b=16\label{subfig-3:exp_search}]{%
      \includegraphics[width=2.3in]{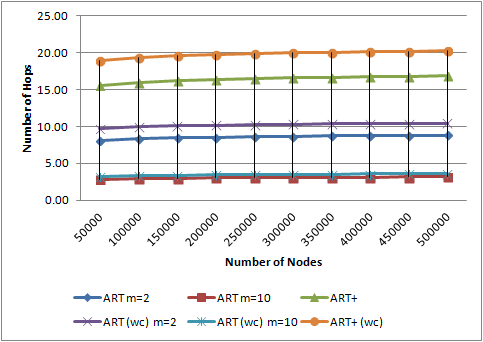}
    }
    \hfill
    \subfloat[massive failures\label{subfig-4:exp_search}]{%
      \includegraphics[width=2.3in]{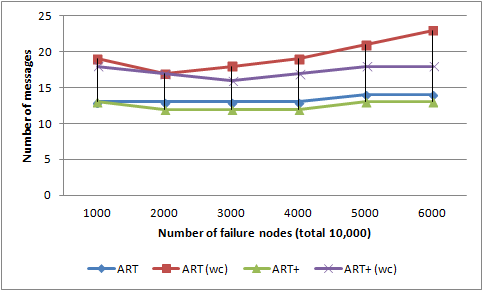}
    }
    \caption{Cost of search operations}
    \label{fig:exp_search}
  \end{figure}

To measure the network performance for the search operations (single and range queries), we conducted experiments for different values of $b$, 2, 4 and 16, in which for each $N$, we executed 1,000 single queries and 1,000 range queries. The search cost is depicted in Fig. \ref{fig:exp_search}. Both normal (beta, uniform) and worst cases (powlow, zipfian, weibull) are depicted in the same graph.
Experiments confirm that the query performance of ART and ART$^+$ is $O(\log_{b}^2{\log{N}})$ and the slight performance divergences are due to the fact that BATON$^*$, as the inner structure of ART's cluster-peer, performs better that D$^3$-Tree in search operations.

In case of massive failures, the search algorithm has to find alternative paths to overcome the unreachable peers. Thus, an increase in node failures results in an increase in search costs. To evaluate the system in case of massive failures, we initialized the system with 10,000 peers and let them randomly fail without recovering. At each step, we check if the network is partitioned or not. Since the backbone of ART and ART$^+$ remains unaffected w.h.p., the search cost is restricted inside a cluster-peer (BATON$^*$ or D$^3$-Tree respectively), meaning that $b$ parameter does not affect the overall expected cost. Figure \ref{subfig-4:exp_search} illustrates the effect of massive failures.
We observe that both structures are fault tolerant since the failure percentage has to reach the threshold of $60\%$ to partition them. Moreover, even in the worst case scenario, the ART$^+$ maintains lower search cost compared to ART, since D$^3$-Tree handles node failures more effectively than BATON$^*$.

\subsubsection{Cost of Load-Balancing Operations.}

\begin{figure}[]
\centering
	\includegraphics[width=2.3in]{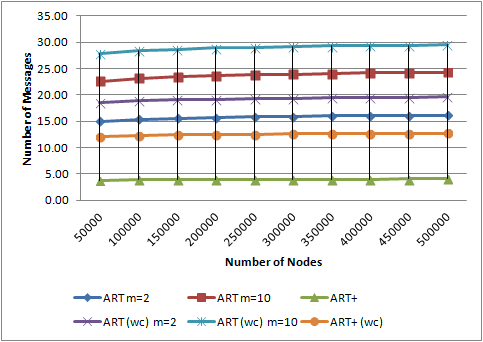}
    \caption{Cost of load-balancing operation}
    \label{fig:exp_loadbalancing}
  \end{figure}

To evaluate the cost of load-balancing, we tested the network with a variety of distributions. For a network of $N$ total nodes, we performed $2N$ node joins. Both ART and ART$^+$ remain unaffected w.h.p., when peers join or leave the network, thus the load-balancing performance is restricted inside a cluster-peer (BATON$^*$ or D$^3$-Tree respectively), meaning that $b$ parameter does not affect the overall expected cost. The load-balancing cost is depicted in Fig. \ref{fig:exp_loadbalancing}. Both normal and worst cases are depicted in the same graph.

Experiments confirm that ART$^+$ has an $O(\log{\log{N}})$ load-balancing performance,  instead of the ART performance of $O(m \cdot \log_m{\log{N}})$. Thus, even in the worst case scenario, the ART$^+$ outperforms ART, since D$^3$-Tree has a more efficient load-balancing mechanism than BATON$^*$ (Fig. \ref{fig:exp_loadbalancing}).
  
\section{Conclusions}
\label{sec:conclusions}

In this work, we presented a dynamic distributed deterministic structure, called D$^3$-Tree. Our proposed structure introduces many enhancements over the solutions in literature and its predecessor, regarding load-balancing and fault tolerance. We presented in brief the theoretical algorithmic analysis, in which D$^3$-Tree is based on, and we described thoroughly the key aspects of the implementation. Verifying the theory, we have proved through experiments that D$^3$-Tree outperforms other well-known tree-based structures, by achieving an $O(\log{N})$ amortized bound in the most costly operation of load-balancing, even in a worst case scenario.
Moreover, investigating the structure's fault tolerance, both theoretically and through experiments, we proved that D$^3$-Tree is highly fault tolerant, since, even for massive node failures of $30\%$, it achieves a significant success rate of 85$\%$ in element search, without increasing the cost considerably.

Afterwards we went one step further, in order to achieve sub-logarithmic complexity and proposed the ART$^+$ structure, exploiting the excellent performance of D$^3$-Tree.
We proved that the communication cost of query operations, element update and node join/leave operations of ART$^+$ scale sub-logarithmically expected w.h.p. Moreover, the cost for the load-balancing operation is sub-logarithmic amortized. Experimental comparison to its predecessor, ART, showed slightly less efficiency towards search operations (single and range queries), but improved performance for the load-balancing operation and the search operations in case of node failures. Moreover, experiments confirmed that ART$^+$ is highly fault tolerant in case of massive failures. Note that, so far, ART outperforms the state-of-the-art decentralized structures.

\subsubsection*{Acknowledgments.} This research has been co-financed by the European Union (European Social Fund - ESF) and Greek national funds through the Operational Program "Education and Lifelong Learning" of the National Strategic Reference Framework (NSRF) - Research Funding Program: Heracleitus II. Investing in knowledge society through the European Social Fund.

\bibliographystyle{splncs03}
\bibliography{bibliography}

\begin{thebibliography}{10}
\providecommand{\url}[1]{\texttt{#1}}
\providecommand{\urlprefix}{URL }

\bibitem{bstz14}
Brodal, G., Sioutas, S., Tsichlas, K., Zaroliagis, C.: D$^2$-tree: A new
  overlay with deterministic bounds. Algorithmica pp. 1--22 (April 2014)

\bibitem{clmgs2011:pring}
Crainiceanu, A., Linga, P., Machanavajjhala, A., Gehrke, J., Shanmugasundaram,
  J.: Load balancing and range queries in p2p systems using p-ring. ACM Trans.
  Internet Technol.  10(4),  16:1--16:30 (Mar 2011)

\bibitem{GAA2003}
Gupta, A., Agrawal, D., Abbadi, A.E.: Approximate range selection queries in
  peer-to-peer systems. In: In Proc. of the 1st Biennial Conference on
  Innovative Data Systems Research (CIDR 2003) (2003)

\bibitem{jotvz06}
Jagadish, H.V., Ooi, B.C., Tan, K., Vu, Q.H., Zhang, R.: Speeding up search in
  p2p networks with a multi-way tree structure. In: Proc. of ACM International
  Conference on Management of Data (SIGMOD 2006). pp. 1--12. Chicago, Illinois,
  USA (2006)

\bibitem{jov05}
Jagadish, H.V., Ooi, B.C., Vu, Q.H.: Baton: a balanced tree structure for
  peer-to-peer networks. In: Proc. of the 31st Conference on Very Large
  Databases (VLDB '05). pp. 661--672. Trondheim, Norway (2005)

\bibitem{kmsttz2003}
Kaporis, A., Makris, C., Sioutas, S., Tsakalidis, A., Tsichlas, K., Zaroliagis,
  C.: Improved bounds for finger search on a ram. In: Proc. of 11th Annual
  European Symposium on Algorithms (ESA). vol. 2832, pp. 325--336 (September
  2003)

\bibitem{ov2011}
Ozsu, M.T., Valduriez, P.: Principles of Distributed Database Systems. Springer
  (2011)

\bibitem{SGAA2004}
Sahin, O., Gupta, A., Agrawal, D., Abbadi, A.E.: A peer-to-peer framework for
  caching range queries. In: Proc. of the 20th Conference on Data Engineering
  (ICDE 2004). pp. 165--176. IEEE (March 2004)

\bibitem{s02:p2pdef}
Schollmeier, R.: A definition of peer-to-peer networking for the classification
  of peer-to-peer architectures and applications. In: Proc. of 1st
  International Conference on Peer-to-Peer Computing. IEEE (2002)

\bibitem{spstm:p2psim}
Sioutas, S., Papaloukopoulos, G., Sakkopoulos, E., Tsichlas, K., Manolopoulos,
  Y.: A novel distributed p2p simulator architecture: D-p2p-sim. In: ACM CIKM.
  pp. 2069--2070 (2009)

\bibitem{stpstm2012:art}
Sioutas, S., Triantafillou, P., Papaloukopoulos, G., Sakkopoulos, E., Tsichlas,
  K.: Art: Sub-logarithmic decentralized range query processing with
  probabilistic guarantees. Journal of Distributed and Parallel Databases
  (DAPD)  31(1),  71--109 (2012)

\bibitem{Stoica:chord}
Stoica, I., Morris, R., Karger, D., Kaashoek, M.F., Balakrishnan, H.: Chord: A
  scalable peer-to-peer lookup service for internet applications. SIGCOMM
  Comput. Commun. Rev.  31(4),  149--160 (Aug 2001)

\end{thebibliography}

\end{document}